\documentclass[a4paper,10pt,twoside]{article}
\pdfoutput=1
\usepackage{newtxtext}
\usepackage{newtxmath}
\usepackage[zerostyle=c]{newtxtt}
\usepackage[T1]{fontenc}
\usepackage{amsmath}
\usepackage{bm}
\usepackage[hidelinks]{hyperref}
\usepackage{graphicx}
\usepackage[top=3cm,bottom=2.5cm,left=2.5cm,right=2.5cm]{geometry}
\usepackage[labelfont={bf},labelsep=period,justification=justified,hang]{caption}
\usepackage[caption=false,font=footnotesize]{subfig}
\usepackage{xcolor}
\usepackage{ifthen}
\usepackage{fancyhdr}
\usepackage{url}
\usepackage{eso-pic}
\usepackage{etoolbox}
\usepackage{siunitx}
\usepackage{trfsigns}
\usepackage{enotez}
\setenotez{list-name=Errata,counter-format=alph}

\graphicspath{{./gfx/}}
\newcommand{\Title}[1]{\def\Title{#1}}
\newcommand{\TitleRunning}[1]{\def\TitleRunning{#1}}
\newcommand{\Author}[1]{\def\Author{#1}}
\newcommand{\AuthorRunning}[1]{\def\AuthorRunning{#1}}
\newcommand{\Address}[1]{\def\Address{#1}}
\newcommand{\Abstract}[1]{\def\Abstract{#1}}

\renewcommand{\maketitle}{%
    \thispagestyle{empty}%
    \begin{center}
    \ \\[1ex]
    {\Large\bfseries\Title}
    \\[1em]
    {\large\Author}
    \\[0.5em]
    {\Address}
    \\[1em]
    \parbox{0.8\linewidth}{\paragraph*{Abstract}\Abstract}
    \end{center}
    \vspace{1em}
}

\newcommand{\D}{\displaystyle}
\newcommand{\Vector}[1]{\bm{#1}}  
\newcommand{\Matrix}[1]{\bm{#1}}  
\newcommand{\Person}[1]{#1}
\newcommand{\refEq}[1]{(\ref{#1})}               
\newcommand{\refFigBegin}[1]{Figure~\ref{#1}}    
\newcommand{\refSec}[1]{Section~\ref{#1}}        

\Title{A Simulative Study on Active Disturbance Rejection Control (ADRC)\protect\\ as a Control Tool for Practitioners}
\TitleRunning{A Simulative Study on Active Disturbance Rejection Control (ADRC) as a Control Tool for Practitioners}

\AuthorRunning{Gernot Herbst}%
\Author{Gernot Herbst%
    \ \href{https://orcid.org/0000-0002-4638-5378}%
    {\raisebox{-0.3pt}{\includegraphics[height=9pt]{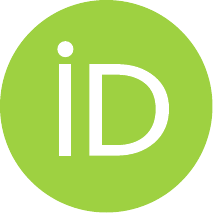}}}%
}
\Address{%
    Siemens AG, Chemnitz, Germany\\%
    \url{gernot.herbst@siemens.com}
}%

\Abstract{%
As an alternative to both classical PID-type and modern model-based approaches to solving control problems, active disturbance rejection control (ADRC) has gained significant traction in recent years. With its simple tuning method and robustness against process parameter variations, it puts itself forward as a valuable addition to the toolbox of control engineering practitioners. This article aims at providing a single-source introduction and reference to linear ADRC with this audience in mind. A simulative study is carried out using generic first- and second-order plants to enable a quick visual assessment of the abilities of ADRC. Finally, a modified form of the discrete-time case is introduced to speed up real-time implementations as necessary in applications with high dynamic requirements.
}

\hypersetup{%
    pdfauthor={\AuthorRunning},
    pdftitle={\TitleRunning},
    pdfcreator={},pdfproducer={}
}

\begin{document}

\AddToShipoutPicture*{\AtPageUpperLeft{\put(\LenToUnit{2cm},\LenToUnit{-1.6cm}){%
   \color{black!50}%
   \begin{minipage}[t]{17cm}
   This is a preprint version.
   The final article is available at \textcolor{blue!60}{\href{http://dx.doi.org/10.3390/electronics2030246}{doi:10.3390/electronics2030246}}.
   Please cite as:\\
   G. Herbst, ``A Simulative Study on Active Disturbance Rejection Control (ADRC) as a Control Tool for Practitioners,'' \emph{Electronics}, vol. 2, no. 3, pp. 246--279, Aug. 2013.\\
   This revision of the preprint includes post-publication corrections.
   \end{minipage}
}}}

\maketitle


\section{Introduction}

Active disturbance rejection control (ADRC) \cite{Han:2009, Gao:2001, Gao:2003, Gao:2006} has emerged as an alternative that combines easy applicability known from classical PID-type control methods with the power of modern model-based approaches. The foundation for ADRC is an observer who jointly treats actual disturbances and modeling uncertainties, such that only a very coarse process model is necessary in order to design a control loop, which makes ADRC an attractive choice for practitioners and promises good robustness against process variations. Present applications range from power electronics \cite{Sun:2005}, motion control \cite{Su:2005} and superconducting radio frequency cavities \cite{Vincent:2011} to tension and temperature control \cite{Zheng:2010}.

While it can be shown (\textit{cf}.\ Section \ref{sec:adrc_relation}) that the linear case of ADRC is equivalent to a special case of classical state space control with disturbance estimation and compensation based on the internal model principle \cite{Francis:1976}, there is an important difference to model-based approaches, such as model predictive control \cite{Qin:2003} or embedded model control \cite{Canuto:2007}: for the latter, an explicit model of the process to be controlled is necessary. ADRC, on the other hand, does only assume a certain canonical model regardless of the actual process dynamics and leaves all modeling errors to be handled as a disturbance. Of course this may come at the price of performance losses compared to a controller built around a precise process model or a model of the reference trajectory. Therefore, while employing the same mathematical tools, ADRC's unified view and treatment of disturbances can be seen as a certain departure from the model-based control school \cite{Radke:2006}, shifting back the focus from modeling to control. This may be key to its appeal for practitioners.

The remainder of this article is organized as follows: After providing a step-by-step introduction to the linear case of ADRC in the following section, a series of simulative experiments is carried out to demonstrate the abilities of ADRC when being faced with varying process parameters or structural uncertainties and to visually provide insights into the effect of its tuning parameters. For the discrete-time case, which is introduced afterwards, an optimized formulation is presented, which enables a controller implementation with very low input-output delay.


\section{Linear Active Disturbance Rejection Control}
\label{sec:adrc}

The aim of this section is to repeat and present the linear case of ADRC in a self-contained manner, following \cite{Gao:2003,Chen:2011}. While the majority of articles introduces ADRC with a second-order process, here the first-order case will be considered first and explicitly, due to its practical importance, since there are many systems---albeit technically nonlinear and higher-order---which exhibit a dominating first-order-like behavior, at least in certain operating points. The second-order case will be developed subsequently with a similar use case in mind. Additionally, it will be shown in Section \ref{sec:adrc_relation} that linear ADRC can be seen as a special case from the perspective of classical state space control with disturbance compensation based on the internal model principle.


\subsection{First-Order ADRC}
\label{sec:adrc1}

Consider a simple first-order process, $P(s)$, with a DC gain, $K$, and a time constant, $T$:
\begin{equation}
\label{eqn:pt1}
P(s) = \D\frac{y(s)}{u(s)} = \D\frac{K}{Ts + 1}
\quad \Laplace \quad
T \cdot \dot{y}(t) + y(t) = K \cdot u(t)
\end{equation}

We add an input disturbance, $d(t)$, to the process, abbreviate $b = \frac{K}{T}$ and rearrange:
\begin{equation*}
\dot{y}(t)
= -\frac{1}{T} \cdot y(t) + \frac{1}{T} \cdot d(t) + \frac{K}{T} \cdot u(t)
= -\frac{1}{T} \cdot y(t) + \frac{1}{T} \cdot d(t) + b \cdot u(t)
\end{equation*}

As our last modeling step, we substitute $b = b_0 + \Delta b$, where $b_0$ shall represent the known part of $b = \frac{K}{T}$ and $\Delta b$, an (unknown) modeling error, and, finally, obtain Equation \refEq{eqn:pt1_with_disturbance}. We will see soon that all that we need to know about our first-order process to design an ADRC is $b_0 \approx b$, \textit{i.e}., an approximate value of $\frac{K}{T}$. Modeling errors or varying process parameters are represented by $\Delta b$ and will be handled~internally.
\begin{equation}
\label{eqn:pt1_with_disturbance}
\dot{y}(t)
= \underbrace{ \left( -\frac{1}{T} \cdot y(t) + \frac{1}{T} \cdot d(t) + \Delta b \cdot u(t) \right) }_{\text{generalized disturbance} f(t)} + b_0 \cdot u(t)
= f(t) + b_0 \cdot u(t)
\end{equation}

By combining $-\frac{1}{T} \cdot y(t)$, the disturbance $d(t)$, and the unknown part $\Delta b \cdot u(t)$ to a so-called generalized disturbance, $f(t)$, the model for our process changed from a first-order low-pass type to an integrator. The fundamental idea of ADRC is to implement an extended state observer (ESO) that provides an estimate, $\hat{f}(t)$, such that we can compensate the impact of $f(t)$ on our process (model) by means of disturbance rejection. All that remains to be handled by the actual controller will then be a process with approximately integrating behavior, which can easily be done, e.g.\ by means of a simple proportional controller.

In order to derive the estimator, a state space description of the disturbed process in Equation \refEq{eqn:pt1_with_disturbance} is~necessary:
\begin{equation}
\label{eqn:pt1_with_disturbance_ss}
\begin{split}
\begin{pmatrix}
\dot{x}_1(t) \\ \dot{x}_2(t)
\end{pmatrix}
&=
\underbrace{
\begin{pmatrix}
0 & 1 \\
0 & 0
\end{pmatrix}
}_{\Matrix{A}}
\cdot
\begin{pmatrix}
x_1(t) \\ x_2(t)
\end{pmatrix}
+
\underbrace{
\begin{pmatrix}
b_0 \\ 0
\end{pmatrix}
}_{\Matrix{B}}
\cdot
u(t)
+
\begin{pmatrix}
0 \\ 1
\end{pmatrix}
\cdot
\dot{f}(t)
\\
y(t) &=
\underbrace{
\begin{pmatrix}
1 & 0
\end{pmatrix}
}_{\Matrix{C}}
\cdot
\begin{pmatrix}
x_1(t) \\ x_2(t)
\end{pmatrix}
\end{split}
\end{equation}

Since the ``virtual'' input, $\dot{f}(t)$, cannot be measured, a state observer for this kind of process can, of course, only be built using the input, $u(t)$, and output, $y(t)$, of the process. An estimated state, $\hat{x}_2(t)$, however, will provide an approximate value of $f(t)$, \textit{i.e}., $\hat{f}(t)$, if the actual generalized disturbance, $f(t)$, can be considered piecewise constant. The equations for the extended state observer (integrator process extended by a generalized disturbance) are given in Equation \refEq{eqn:pt1_eso}. Note that for linear ADRC, a Luenberger observer is being used, while in the original case of ADRC, a nonlinear observer was employed \cite{Gao:2006}.
\begin{equation}
\label{eqn:pt1_eso}
\begin{split}
\begin{pmatrix}
\dot{\hat{x}}_1(t) \\ \dot{\hat{x}}_2(t)
\end{pmatrix}
&=
\begin{pmatrix}
0 & 1 \\
0 & 0
\end{pmatrix}
\cdot
\begin{pmatrix}
\hat{x}_1(t) \\ \hat{x}_2(t)
\end{pmatrix}
+
\begin{pmatrix}
b_0 \\ 0
\end{pmatrix}
\cdot
u(t)
+
\begin{pmatrix}
l_1 \\ l_2
\end{pmatrix}
\cdot
\left( y(t) - \hat{x}_1(t) \right)
\\
&=
\underbrace{
\begin{pmatrix}
-l_1 & 1 \\
-l_2 & 0
\end{pmatrix}
}_{\Matrix{A} - \Matrix{L}\Matrix{C}}
\cdot
\begin{pmatrix}
\hat{x}_1(t) \\ \hat{x}_2(t)
\end{pmatrix}
+
\underbrace{
\begin{pmatrix}
b_0 \\ 0
\end{pmatrix}
}_{\Matrix{B}}
\cdot
u(t)
+
\underbrace{
\begin{pmatrix}
l_1 \\ l_2
\end{pmatrix}
}_{\Matrix{L}}
\cdot
y(t)
\end{split}
\end{equation}

One can now use the estimated variables, $\hat{x}_1(t) = \hat{y}(t)$ and $\hat{x}_2(t) = \hat{f}(t)$, to implement the disturbance rejection and the actual controller.
\begin{equation}
\label{eqn:adrc1_controller}
u(t) = \D\frac{u_0(t) - \hat{f}(t)}{b_0}
\quad \text{with} \quad
u_0(t) = K_\mathrm{P} \cdot \left( r(t) - \hat{y}(t) \right)
\end{equation}

\begin{figure}[t]
 \centering%
 \includegraphics{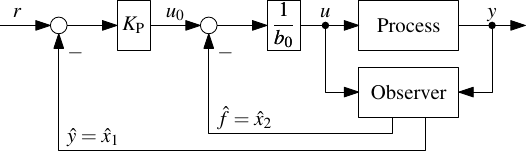}%
 \caption{Control loop structure with ADRC for a first-order process.}
 \label{fig:adrc1_structure}
\end{figure}

The according structure of the control loop is presented in Figure \ref{fig:adrc1_structure}. Since $K_\mathrm{P}$ acts on $\hat{y}(t)$, rather than the actual output $y(t)$, we do have a estimation-based state feedback controller, but the resemblance to a classical proportional controller is striking to practitioners. In Equation \refEq{eqn:adrc1_controller}, $u_0(t)$ represents the output of a linear proportional controller. The remainder of the control law in $u(t)$ is chosen such that the linear controller acts on a normalized integrator process if $\hat{f}(t) \approx f(t)$ holds. The effect can be seen by putting Equation \refEq{eqn:adrc1_controller} in Equation \refEq{eqn:pt1_with_disturbance}:
\begin{equation*}
\dot{y}(t)
= f(t) + b_0 \cdot \D\frac{u_0(t) - \hat{f}(t)}{b_0}
= \left( f(t) - \hat{f}(t) \right) + u_0(t)
\approx u_0(t)
= K_\mathrm{P} \cdot \left( r(t) - \hat{y}(t) \right)
\end{equation*}

If $\hat{y}(t) \approx y(t)$ holds, we obtain a first-order closed loop behavior with a pole, $s^\mathrm{CL} = -K_\mathrm{P}$:
\begin{equation*}
\D\frac{1}{K_\mathrm{P}} \cdot \dot{y}(t) + \hat{y}(t)
\approx \D\frac{1}{K_\mathrm{P}} \cdot \dot{y}(t) + y(t)
\approx r(t)
\end{equation*}

If the state estimator and disturbance rejection work properly, one has to design a proportional controller only one single time to obtain the same closed loop behavior, regardless of the parameters of the actual process. For example, one can calculate $K_\mathrm{P}$ from a desired first-order system with 2\%-settling~time:
\begin{equation}
\label{eqn:adrc1_kp}
K_\mathrm{P} \approx \D\frac{4}{T_\mathrm{settle}}
\end{equation}

In order to work properly, observer parameters, $l_1$ and $l_2$, in Equation \refEq{eqn:pt1_eso} still have to be determined. Since the observer dynamics must be fast enough, the observer poles, $s^\mathrm{ESO}_{1/2}$, must be placed left of the closed loop pole, $s^\mathrm{CL}$. A simple rule of thumb suggests for both poles:
\begin{equation}
\label{eqn:adrc1_pole_eso}
s^\mathrm{ESO}_{1/2} = s^\mathrm{ESO} \approx (3 \ldots 10) \cdot s^\mathrm{CL}
\quad \text{with} \quad
s^\mathrm{CL} = -K_\mathrm{P} \approx -\D\frac{4}{T_\mathrm{settle}}
\end{equation}

Placing all observer poles at one location is also known as ``bandwidth parameterization'' \cite{Gao:2003}. Since the matrix $(\Matrix{A} - \Matrix{L}\Matrix{C})$ determines the error dynamics of the observer, we can compute the necessary observer gains for the common pole location, $s^\mathrm{ESO}$, from its characteristic polynomial:
\begin{equation*}
\det\left( s\Matrix{I} - \left(\Matrix{A} - \Matrix{L}\Matrix{C}\right)\right)
= s^2 + l_1 \cdot s + l_2
\stackrel{!}{=} \left(s - s^\mathrm{ESO}\right)^2 = s^2 - 2 s^\mathrm{ESO} \cdot s + \left(s^\mathrm{ESO}\right)^2
\end{equation*}

From this equation, the solutions for $l_1$ and $l_2$ can immediately be read off:
\begin{equation}
\label{eqn:adrc1_l_eso}
l_1 = -2 \cdot s^\mathrm{ESO} \quad \text{and} \quad l_2 = \left(s^\mathrm{ESO}\right)^2
\end{equation}

To summarize, in order to implement a linear ADRC for a first-order system, four steps are necessary:
\begin{enumerate}
\item
\emph{Modeling:}
For a process with (dominating) first-order behavior, 
$P(s) = \D\frac{K}{Ts + 1}$, all that needs to be known is an estimate $b_0 \approx \frac{K}{T}$.

\item
\emph{Control structure:}
Implement a proportional controller with disturbance rejection and an extended state observer, as given in Equations \refEq{eqn:pt1_eso} and \refEq{eqn:adrc1_controller}:

\medskip
$
\begin{pmatrix}
\dot{\hat{x}}_1(t) \\ \dot{\hat{x}}_2(t)
\end{pmatrix}
=
\begin{pmatrix}
-l_1 & 1 \\
-l_2 & 0
\end{pmatrix}
\cdot
\begin{pmatrix}
\hat{x}_1(t) \\ \hat{x}_2(t)
\end{pmatrix}
+
\begin{pmatrix}
b_0 \\ 0
\end{pmatrix}
\cdot
u(t)
+
\begin{pmatrix}
l_1 \\ l_2
\end{pmatrix}
\cdot
y(t)
$

\medskip
$u(t)
= \D\frac{K_\mathrm{P} \cdot \left( r(t) - \hat{y}(t) \right) - \hat{f}(t)}{b_0}
= \frac{K_\mathrm{P} \cdot \left( r(t) - \hat{x}_1(t) \right) - \hat{x}_2(t)}{b_0}
$
\medskip

\item
\emph{Closed loop dynamics:}
Choose $K_\mathrm{P}$, e.g.\ according to a desired settling time Equation \refEq{eqn:adrc1_kp}:\\
$K_\mathrm{P} \approx \D\frac{4}{T_\mathrm{settle}}$.
\medskip

\item
\emph{Observer dynamics:}
Place the observer poles left of the closed loop pole via Equations \refEq{eqn:adrc1_pole_eso} and \refEq{eqn:adrc1_l_eso}:

$l_1 = -2 \cdot s^\mathrm{ESO}$, \quad $l_2 = \left(s^\mathrm{ESO}\right)^2$
\quad with \quad
$s^\mathrm{ESO} \approx (3 \ldots 10) \cdot s^\mathrm{CL}$
\quad and \quad
$s^\mathrm{CL} = -K_\mathrm{P}$

\end{enumerate}

It should be noted that the same control structure can be applied to a first-order integrating process:
\begin{equation*}
P(s) = \D\frac{y(s)}{u(s)} = \D\frac{K_\mathrm{I}}{s}
\quad \Laplace \quad
y(t) = K_\mathrm{I} \cdot u(t)
\end{equation*}

With an input disturbance, $d(t)$, and a substitution, $K_\mathrm{I} = b = b_0 + \Delta b$, with $\Delta b$ representing the unknown part of $K_\mathrm{I}$, we can model the process in an identical manner as Equation \refEq{eqn:pt1_with_disturbance}, with all differences hidden in the generalized disturbance, $f(t)$:
\begin{equation*}
\dot{y}(t)
= \underbrace{ \left( d(t) + \Delta b \cdot u(t) \right) }_{\text{gen.\ disturbance} f(t)} + b_0 \cdot u(t)
= f(t) + b_0 \cdot u(t)
\end{equation*}

Therefore, the design of the ADRC for a first-order integrating process can follow the same four design steps given above, with the only distinction that $b_0$ must be set to $b_0 \approx K_\mathrm{I}$ in step~1.


\subsection{Second-Order ADRC}
\label{sec:adrc2}

Following the previous section, we now consider a second-order process, $P(s)$, with a DC gain, $K$, damping factor, $D$, and a time constant, $T$.
\begin{equation}
\label{eqn:pt2}
P(s) = \D\frac{y(s)}{u(s)} = \D\frac{K}{T^2 s^2 + 2DTs + 1}
\quad \Laplace \quad
T^2 \cdot \ddot{y}(t) + 2DT \cdot \ddot{y}(t) + y(t) = K \cdot u(t)
\end{equation}

As for the first-order case, we add an input disturbance, $d(t)$, abbreviate $b = \frac{K}{T^2}$ and split $b$ into a known and unknown part, $b = b_0 + \Delta b$:
\begin{equation}
\label{eqn:pt2_with_disturbance}
\ddot{y}(t)
= \underbrace{ \left( -\frac{2D}{T} \cdot \dot{y}(t) - \frac{1}{T^2} \cdot y(t) + \frac{1}{T^2} \cdot d(t) + \Delta b \cdot u(t) \right) }_{\text{generalized disturbance} f(t)} + b_0 \cdot u(t)
= f(t) + b_0 \cdot u(t)
\end{equation}

With everything else combined into the generalized disturbance, $f(t)$, all that remains of the process model is a double integrator. The state space representation of the disturbed double integrator is:
\begin{equation}
\label{eqn:pt2_with_disturbance_ss}
\begin{split}
\begin{pmatrix}
\dot{x}_1(t) \\ \dot{x}_2(t) \\ \dot{x}_3(t)
\end{pmatrix}
&=
\underbrace{
\begin{pmatrix}
0 & 1 & 0 \\
0 & 0 & 1 \\
0 & 0 & 0 \\
\end{pmatrix}
}_{\Matrix{A}}
\cdot
\begin{pmatrix}
x_1(t) \\ x_2(t) \\ x_3(t)
\end{pmatrix}
+
\underbrace{
\begin{pmatrix}
0 \\ b_0 \\ 0
\end{pmatrix}
}_{\Matrix{B}}
\cdot
u(t)
+
\begin{pmatrix}
0 \\ 0 \\ 1
\end{pmatrix}
\cdot
\dot{f}(t)
\\
y(t) &=
\underbrace{
\begin{pmatrix}
1 & 0 & 0
\end{pmatrix}
}_{\Matrix{C}}
\cdot
\begin{pmatrix}
x_1(t) \\ x_2(t) \\ x_3(t)
\end{pmatrix}
\end{split}
\end{equation}

In order to employ a control law similar to the first-order case, an extended state observer is needed to provide an estimation, $\hat{x}_1(t) = \hat{y}(t)$, $\hat{x}_2(t) = \dot{\hat{y}}(t)$ and $\hat{x}_3(t) = \hat{f}(t)$:
\begin{equation}
\label{eqn:pt2_eso}
\begin{split}
\begin{pmatrix}
\dot{\hat{x}}_1(t) \\ \dot{\hat{x}}_2(t) \\ \dot{\hat{x}}_3(t)
\end{pmatrix}
&=
\begin{pmatrix}
0 & 1 & 0 \\
0 & 0 & 1 \\
0 & 0 & 0 \\
\end{pmatrix}
\cdot
\begin{pmatrix}
\hat{x}_1(t) \\ \hat{x}_2(t) \\ \hat{x}_3(t)
\end{pmatrix}
+
\begin{pmatrix}
0 \\ b_0 \\ 0
\end{pmatrix}
\cdot
u(t)
+
\begin{pmatrix}
l_1 \\ l_2 \\ l_3
\end{pmatrix}
\cdot
\left( y(t) - \hat{x}_1(t) \right)
\\
&=
\underbrace{
\begin{pmatrix}
-l_1 & 1 & 0 \\
-l_2 & 0 & 1 \\
-l_3 & 0 & 0 \\
\end{pmatrix}
}_{\Matrix{A} - \Matrix{L}\Matrix{C}}
\cdot
\begin{pmatrix}
\hat{x}_1(t) \\ \hat{x}_2(t) \\ \hat{x}_3(t)
\end{pmatrix}
+
\underbrace{
\begin{pmatrix}
0 \\ b_0 \\ 0
\end{pmatrix}
}_{\Matrix{B}}
\cdot
u(t)
+
\underbrace{
\begin{pmatrix}
l_1 \\ l_2 \\ l_3
\end{pmatrix}
}_{\Matrix{L}}
\cdot
y(t)
\end{split}
\end{equation}

\begin{figure}
 \centering%
 \includegraphics{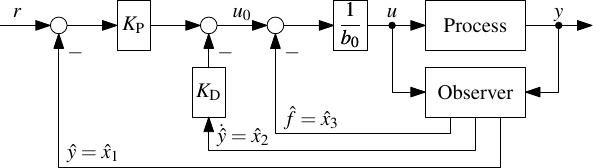}%
 \caption{Control loop structure with active disturbance rejection control (ADRC) for a second-order process.}
 \label{fig:adrc2_structure}
\end{figure}

Using the estimated variables, one can implement the disturbance rejection and a linear controller for the remaining double integrator behavior, as shown in Figure \ref{fig:adrc2_structure}. A modified PD controller (without the derivative part for the reference value $r(t)$) will lead to a second-order closed loop behavior with adjustable dynamics. Again, this actually is an estimation-based state feedback controller.
\begin{equation}
\label{eqn:adrc2_controller}
u(t) = \D\frac{u_0(t) - \hat{f}(t)}{b_0}
\quad \text{with} \quad
u_0(t) = K_\mathrm{P} \cdot \left( r(t) - \hat{y}(t) \right) - K_\mathrm{D} \cdot \dot{\hat{y}}(t)
\end{equation}

Provided the estimator delivers good estimates, $\hat{x}_1(t) = \hat{y}(t) \approx y(t)$, $\hat{x}_2(t) = \dot{\hat{y}}(t) \approx \dot{y}(t)$ and $\hat{x}_3(t) = \hat{f}(t) \approx f(t)$, one obtains after inserting Equation \refEq{eqn:adrc2_controller} into Equation \refEq{eqn:pt2_with_disturbance}:
\begin{equation*}
\ddot{y}(t)
= \left( f(t) - \hat{f}(t) \right) + u_0(t)
\approx u_0(t)
\approx K_\mathrm{P} \cdot \left( r(t) - {y}(t) \right) - K_\mathrm{D} \cdot \dot{y}(t)
\end{equation*}

Under ideal conditions, this leads to:
\begin{equation*}
\D\frac{1}{K_\mathrm{P}} \cdot \ddot{y}(t) + \frac{K_\mathrm{D}}{K_\mathrm{P}} \cdot \dot{y}(t) + y(t)
= r(t)
\end{equation*}

While any second-order dynamics can be set using $K_\mathrm{P}$ and $K_\mathrm{D}$, one practical approach is to tune the closed loop to a critically damped behavior and a desired 2\% settling time $T_\mathrm{settle}$, \textit{i.e}., choose $K_\mathrm{P}$ and $K_\mathrm{D}$ to get a negative-real double pole, $s^\mathrm{CL}_{1/2} = s^\mathrm{CL}$:
\begin{equation}
\label{eqn:adrc2_kp_kd}
K_\mathrm{P} = \left(s^\mathrm{CL}\right)^2
\quad \text{and} \quad
K_\mathrm{D} = -2 \cdot s^\mathrm{CL}
\quad \text{with} \quad
s^\mathrm{CL} \approx -\D\frac{6}{T_\mathrm{settle}}
\end{equation}

Similar to the first order case, the placement of the observer poles can follow a common rule of thumb:
\begin{equation}
\label{eqn:adrc2_pole_eso}
s^\mathrm{ESO}_{1/2/3} = s^\mathrm{ESO} \approx (3 \ldots 10) \cdot s^\mathrm{CL}
\quad \text{with} \quad
s^\mathrm{CL} \approx -\D\frac{6}{T_\mathrm{settle}}
\end{equation}

Once the pole locations are chosen in this manner, the observer gains are computed from the characteristic polynomial of $(\Matrix{A} - \Matrix{L}\Matrix{C})$:
\begin{equation*}
\det\left( s\Matrix{I} - \left(\Matrix{A} - \Matrix{L}\Matrix{C}\right)\right)
= s^3 + l_1 \cdot s^2 + l_2 \cdot s + l_3
\stackrel{!}{=} \left(s - s^\mathrm{ESO}\right)^3 = s^3 - 3 s^\mathrm{ESO} \cdot s^2 + 3 \cdot \left(s^\mathrm{ESO}\right)^2 \cdot s - \left(s^\mathrm{ESO}\right)^3
\end{equation*}

The respective solutions for $l_1$, $l_2$ and $l_3$ are:
\begin{equation}
\label{eqn:adrc2_l_eso}
l_1 = -3 \cdot s^\mathrm{ESO}, \quad l_2 = 3 \cdot \left(s^\mathrm{ESO}\right)^2 \quad \text{and} \quad l_3 = -\left(s^\mathrm{ESO}\right)^3
\end{equation}

To summarize, ADRC for a second-order system is designed and implemented as follows:
\begin{enumerate}
\item
\emph{Modeling:}
For a process with (dominating) second-order behavior, $P(s) = \D\frac{K}{T^2 s^2 + 2DTs + 1}$, one only needs to know an approximate value $b_0 \approx \frac{K}{T^2}$.

\item
\emph{Control structure:}
Implement a proportional controller with disturbance rejection and an extended state observer, as given in Equations \refEq{eqn:pt2_eso} and \refEq{eqn:adrc2_controller}:

\medskip
$
\begin{pmatrix}
\dot{\hat{x}}_1(t) \\ \dot{\hat{x}}_2(t) \\ \dot{\hat{x}}_3(t)
\end{pmatrix}
=
\begin{pmatrix}
-l_1 & 1 & 0 \\
-l_2 & 0 & 1 \\
-l_3 & 0 & 0 \\
\end{pmatrix}
\cdot
\begin{pmatrix}
\hat{x}_1(t) \\ \hat{x}_2(t) \\ \hat{x}_3(t)
\end{pmatrix}
+
\begin{pmatrix}
0 \\ b_0 \\ 0
\end{pmatrix}
\cdot
u(t)
+
\begin{pmatrix}
l_1 \\ l_2 \\ l_3
\end{pmatrix}
\cdot
y(t)
$

\medskip
$
u(t) = \D\frac{\left( K_\mathrm{P} \cdot \left( r(t) - \hat{y}(t) \right) - K_\mathrm{D} \cdot \dot{\hat{y}}(t) \right) - \hat{f}(t)}{b_0}
= \D\frac{\left( K_\mathrm{P} \cdot \left( r(t) - \hat{x}_1(t) \right) - K_\mathrm{D} \cdot \hat{x}_2(t) \right) - \hat{x}_3(t)}{b_0}
$
\medskip

\item
\emph{Closed loop dynamics:}
Choose $K_\mathrm{P}$ and $K_\mathrm{D}$, e.g.\ according to a desired settling time as given in Equation \refEq{eqn:adrc2_kp_kd}:

$
K_\mathrm{P} = \left(s^\mathrm{CL}\right)^2
,\quad
K_\mathrm{D} = -2 \cdot s^\mathrm{CL}
\quad \text{with} \quad
s^\mathrm{CL} \approx -\D\frac{6}{T_\mathrm{settle}}
$
\medskip

\item
\emph{Observer dynamics:}
Place the observer poles left of the closed loop poles via Equations \refEq{eqn:adrc2_pole_eso} \\
and \refEq{eqn:adrc2_l_eso}:

$
l_1 = -3 \cdot s^\mathrm{ESO}, \quad l_2 = 3 \cdot \left(s^\mathrm{ESO}\right)^2, \quad l_3 = -\left(s^\mathrm{ESO}\right)^3
\quad \text{with} \quad
s^\mathrm{ESO} \approx (3 \ldots 10) \cdot s^\mathrm{CL}
$

\end{enumerate}


\subsection{Relation to Linear State Space Control with Disturbance Estimation and Compensation}
\label{sec:adrc_relation}

Given that linear ADRC only employs tools known from classical linear state space control, how can it be compared to existing approaches? In this section, we will demonstrate that linear ADRC can be related to state space control with disturbance estimation and compensation based on the internal model principle \cite{Francis:1976}. We will start with a linear state space model of a process disturbed by $d(t)$, as follows:
\begin{equation}
\label{eqn:linear_plant}
\Vector{\dot{x}}(t) = \Matrix{A} \Vector{x}(t) + \Matrix{B} \Vector{u}(t) + \Matrix{E} \Vector{d}(t)
,\quad
\Vector{y}(t) = \Matrix{C} \Vector{x}(t)
\end{equation}

Further, we assume to possess a model for the generation of the disturbance, $d(t)$:
\begin{equation}
\label{eqn:linear_disturbance}
\Vector{\dot{x}}_\mathrm{d}(t) = \Matrix{A}_\mathrm{d} \Vector{x}_\mathrm{d}(t)
,\quad
\Vector{d}(t) = \Matrix{C}_\mathrm{d} \Vector{x}_\mathrm{d}(t)
\end{equation}

Note that $\Matrix{A}_\mathrm{d}$ and $\Matrix{C}_\mathrm{d}$ in Equation~\refEq{eqn:linear_disturbance} refer to modeling the disturbance generator in this section only and should not be mistaken for the discrete-time versions of $\Matrix{A}$ and $\Matrix{C}$ used in \refSec{sec:adrc_discrete}. The process model is now being extended by incorporating the internal state variables, $\Vector{x}_\mathrm{d}(t)$, of the disturbance generator, resulting in an augmented process model in Equation \refEq{eqn:linear_augmented}, for which an observer given in Equation \refEq{eqn:linear_observer} can be set up \cite{Ostertag:2011}:
\begin{equation}
\label{eqn:linear_augmented}
\begin{pmatrix}
\Vector{\dot{x}}(t) \\ \Vector{\dot{x}}_\mathrm{d}(t)
\end{pmatrix}
=
\begin{pmatrix}
\Matrix{A} & \Matrix{E} \Matrix{C}_\mathrm{d} \\
0 & \Matrix{A}_\mathrm{d}
\end{pmatrix}
\cdot
\begin{pmatrix}
\Vector{x}(t) \\ \Vector{x}_\mathrm{d}(t)
\end{pmatrix}
+
\begin{pmatrix}
\Matrix{B} \\ \Matrix{0}
\end{pmatrix}
\cdot
\Vector{u}(t)
,\quad
y(t) =
\begin{pmatrix}
\Matrix{C} & 0
\end{pmatrix}
\cdot
\begin{pmatrix}
\Vector{x}(t) \\ \Vector{x}_\mathrm{d}(t)
\end{pmatrix}
\end{equation}

\begin{equation}
\label{eqn:linear_observer}
\begin{pmatrix}
\Vector{\dot{\hat{x}}}(t) \\ \Vector{\dot{\hat{x}}}_\mathrm{d}(t)
\end{pmatrix}
=
\underbrace{
\begin{pmatrix}
\Matrix{A} & \Matrix{E} \Matrix{C}_\mathrm{d} \\
0 & \Matrix{A}_\mathrm{d}
\end{pmatrix}
}_{\Matrix{\widetilde{A}}}
\cdot
\begin{pmatrix}
\Vector{\hat{x}}(t) \\ \Vector{\hat{x}}_\mathrm{d}(t)
\end{pmatrix}
+
\underbrace{
\begin{pmatrix}
\Matrix{B} \\ \Matrix{0}
\end{pmatrix}
}_{\Matrix{\widetilde{B}}}
\cdot
\Vector{u}(t)
+
\underbrace{
\begin{pmatrix}
\Matrix{L} \\ \Matrix{L}_\mathrm{d}
\end{pmatrix}
}_{\Matrix{\widetilde{L}}}
\cdot
\Bigg( y(t) -
\underbrace{
\begin{pmatrix}
\Matrix{C} & 0
\end{pmatrix}
}_{\Matrix{\widetilde{C}}}
\cdot
\begin{pmatrix}
\Vector{\hat{x}}(t) \\ \Vector{\hat{x}}_\mathrm{d}(t)
\end{pmatrix}
\Bigg)
\end{equation}

Accordingly, the standard state space control law can now be enhanced by the estimated state variables of the disturbance generator in order to compensate or minimize the impact of the disturbance on the process, if a suitable feedback matrix, $\Matrix{K}_\mathrm{d}$, can be found:
\begin{equation}
\label{eqn:linear_controller}
\Vector{u}(t) = G \cdot \Vector{r}(t) - \Matrix{K} \cdot \Vector{\hat{x}}(t) - \Matrix{K}_\mathrm{d} \cdot \Vector{\hat{x}}_\mathrm{d}(t)
\end{equation}

After inserting Equation \refEq{eqn:linear_controller} into Equation \refEq{eqn:linear_plant}, it becomes apparent that---provided an accurate estimation, $\Vector{\hat{x}}_\mathrm{d}(t)$, is available---the disturbance may be compensated to the extent that $\Matrix{B} \Matrix{K}_\mathrm{d} = \Matrix{E} \Matrix{C}_\mathrm{d}$ can be satisfied \cite{Ostertag:2011}.

We will now compare the combined state and disturbance observer based on the augmented process model, as well as the control law to linear ADRC presented before. The first- and second-order case will be distinguished by (a) and (b):
\begin{itemize}
\item
\emph{Process model and disturbance generator:}
When comparing Equations \refEq{eqn:linear_observer} to \refEq{eqn:pt1_eso} and \refEq{eqn:pt2_eso}, respectively, one obtains the (double) integrator process with a constant disturbance model. The respective matrices $\Matrix{E}$ can be found using Equations \refEq{eqn:pt1_with_disturbance} and \refEq{eqn:pt2_with_disturbance}:
\begin{itemize}
\item[(a)]
$\Matrix{A} = 0$,\quad $\Matrix{B} = b_0$,\quad $\Matrix{C} = 1$,\quad
$\Matrix{E} = \frac{1}{T}$,\quad
$\Matrix{A}_\mathrm{d} = 0$,\quad $\Matrix{C}_\mathrm{d} = T = \frac{K}{b_0}$

\item[(b)]
$\Matrix{A} = \begin{pmatrix} 0 & 1 \\ 0 & 0 \end{pmatrix}$,\quad
$\Matrix{B} = \begin{pmatrix} 0 \\ b_0 \end{pmatrix}$,\quad
$\Matrix{C} = \begin{pmatrix} 1 & 0 \end{pmatrix}$,\quad
$\Matrix{E} = \begin{pmatrix} 0 \\ \frac{1}{T^2} \end{pmatrix}$,\quad
$\Matrix{A}_\mathrm{d} = 0$,\quad $\Matrix{C}_\mathrm{d} = T^2 = \frac{K}{b_0}$
\end{itemize}

\item
\emph{Control law:}
The comparison of Equations \refEq{eqn:adrc1_controller} and \refEq{eqn:linear_controller} or \refEq{eqn:adrc2_controller} can be made with $\hat{f} = \hat{x}_\mathrm{d}$ being the estimated state of the disturbance generator:
\begin{itemize}
\item[(a)]
$\D\frac{1}{b_0} \cdot \left( K_\mathrm{P} \cdot \left( r(t) - \hat{y}(t) \right) - \hat{f}(t) \right)
= \D\frac{K_\mathrm{P}}{b_0} \cdot r(t) - \frac{K_\mathrm{P}}{b_0} \cdot \hat{x}_1(t) - \frac{1}{b_0} \cdot \hat{f}(t)$

gives:
\quad
$K_\mathrm{d} = \D\frac{1}{b_0}$,\quad
$K = K_1 = \D\frac{K_\mathrm{P}}{b_0}$,\quad
$G = \D\frac{K_\mathrm{P}}{b_0} = K_1$
\medskip

\item[(b)]
$\D\frac{1}{b_0} \cdot \left( K_\mathrm{P} \cdot \left( r(t) - \hat{y}(t) \right) - K_\mathrm{D} \cdot \dot{\hat{y}}(t) - \hat{f}(t) \right)
= \D\frac{K_\mathrm{P}}{b_0} \cdot r(t) - \frac{K_\mathrm{P}}{b_0} \cdot \hat{x}_1(t) - \frac{K_\mathrm{D}}{b_0} \cdot \hat{x}_2(t) - \frac{1}{b_0} \cdot \hat{f}(t)$

gives:
\quad
$K_\mathrm{d} = \D\frac{1}{b_0}$,\quad
$\Matrix{K} = \begin{pmatrix} K_1 & K_2 \end{pmatrix}$ with \
$K_1 = \D\frac{K_\mathrm{P}}{b_0}$,\
$K_2 = \D\frac{K_\mathrm{D}}{b_0}$,\quad
$G = \D\frac{K_\mathrm{P}}{b_0} = K_1$
\end{itemize}

\end{itemize}

One can see that the observer and control law are equivalent in structure for both linear ADRC and a state space approach based on the internal model principle, and the model of the disturbance generator in ADRC could be made more visible by this comparison. If the standard design procedure of a state space observer and controller with disturbance compensation will lead to the same parameter values as ADRC will be verified subsequently by following the necessary steps to design $\Matrix{K}$, $\Matrix{K}_\mathrm{d}$, $\Matrix{G}$ and $\Matrix{\widetilde{L}}$ based on the same design goals used in linear ADRC before:
\begin{itemize}
\item
\emph{Feedback gain $\Matrix{K}$}:
The closed loop dynamics are determined by the eigenvalues of $\left(\Matrix{A} - \Matrix{B} \Matrix{K}\right)$. For ADRC, all poles were placed on one location, $s^\mathrm{CL}$. With this design goal, one obtains for $\Matrix{K}$:
\begin{itemize}
\item[(a)]
$\det\left(s\Matrix{I} - \left(\Matrix{A} - \Matrix{B} \Matrix{K}\right)\right) \stackrel{!}{=} s - s^\mathrm{CL}$
\quad gives \quad
$K_1 = -\D\frac{s^\mathrm{CL}}{b_0}$

\item[(b)]
$\det\left(s\Matrix{I} - \left(\Matrix{A} - \Matrix{B} \Matrix{K}\right)\right) \stackrel{!}{=} \left( s - s^\mathrm{CL} \right)^2$
\quad gives \quad
$K_1 = \D\frac{\left(s^\mathrm{CL}\right)^2}{b_0}$ \quad and \quad $K_2 = -\D\frac{2 s^\mathrm{CL}}{b_0}$
\end{itemize}

\item
\emph{Gain compensation $\Matrix{G}$:}
In order to eliminate steady state tracking errors, $\Matrix{G}$ must be chosen to
$\Matrix{G} = -\left( \Matrix{C} \cdot \left( \Matrix{A} - \Matrix{B} \Matrix{K} \right)^{-1} \cdot \Matrix{B} \right)^{-1}$,
which gives $G = K_1$ for both the first- and second-order case.

\item
\emph{Observer gain $\Matrix{\widetilde{L}}$}:
The dynamics of the observer for the augmented system are determined by placing the eigenvalues of $\left(\Matrix{\widetilde{A}} - \Matrix{\widetilde{L}}\Matrix{\widetilde{C}}\right)$ as desired, which is the identical procedure, as in Equations \refEq{eqn:adrc1_l_eso} and \refEq{eqn:adrc2_l_eso}.

\item
\emph{Disturbance compensation gain $\Matrix{K}_\mathrm{d}$}:
As mentioned above, $\Matrix{K}_\mathrm{d}$ should be chosen to achieve $\Matrix{B} \Matrix{K}_\mathrm{d} \stackrel{!}{=} \Matrix{E} \Matrix{C}_\mathrm{d}$ if possible:
\begin{itemize}
\item[(a)]
$\Matrix{B} \Matrix{K}_\mathrm{d} = b_0 \cdot K_\mathrm{d} \stackrel{!}{=} \Matrix{E} \Matrix{C}_\mathrm{d} = 1$
\quad gives \quad
$K_\mathrm{d} = \D\frac{1}{b_0}$

\item[(b)]
$\Matrix{B} \Matrix{K}_\mathrm{d} =
\begin{pmatrix} 0 \\ b_0 \cdot K_\mathrm{d} \end{pmatrix}
\stackrel{!}{=} \Matrix{E} \Matrix{C}_\mathrm{d}
\begin{pmatrix} 0 \\ 1 \end{pmatrix}$
\quad gives \quad
$K_\mathrm{d} = \D\frac{1}{b_0}$
\end{itemize}

\end{itemize}

Obviously, both designs deliver the same parameters. Based on this comparison, one may view linear ADRC and its controller design as a special case of classical state space control with an observer using a system model augmented by a certain disturbance generator model (following the internal model principle) and disturbance compensation. However, a subtle, but important, distinction has to be made: while the latter relies on a model of the plant (like all modern model-based control approaches), ADRC does always deliberately assume an integrator model for the plant and leaves all modeling errors to be handled by the disturbance estimation. Therefore, ADRC can be applied without accurately modeling the process, which presents a departure from the model-based control school \cite{Radke:2006}.


\section{Simulative Experiments}
\label{sec:experiments}

In the ideal case---with noise-free measurements and unlimited, ideal actuators in the control loop---ADRC would be able to suppress basically all effects of disturbances and parameter variations of the process. In practice, however, we have to live with constraints, such as limited observer dynamics or saturated controller outputs, leading to a compromise when choosing the parameters of the controller.

This section is meant to provide insights into the abilities and limitations of continuous-time ADRC in a visual manner. To that end, the controller---designed once and then left unchanged---will be confronted with a heavily varying process. The influence of the ESO pole placement (relative to the closed loop poles) will also be examined, as well as limitations of the actuator. The experiments are carried out by means of Matlab/Simulink-based simulations.

In Section \ref{sec:adrc_discrete_sim}, further simulations will be performed for the discrete-time implementation of ADRC, also addressing the effect of noise and sampling time.


\subsection{First-Order ADRC with a First-Order Process}
\label{sec:exp_adrc1}

In a first series of experiments, a continuous-time ADRC (as introduced in Section \ref{sec:adrc1}) operating on a first-order system will be examined. The process structure and nominal parameters used to design the controller are:
\begin{equation*}
P(s) = \D\frac{y(s)}{u(s)} = \D\frac{K}{Ts + 1}
\quad \text{with} \quad
K = 1 \quad \text{and} \quad T = 1
\end{equation*}

The ADRC is designed following the four steps described in Section \ref{sec:adrc1}. For now, we assume perfect knowledge of our process and set $b_0 = \frac{K}{T} = 1$, but will leave this value unchanged for the rest of the experiments in this section. The desired closed loop settling time is chosen, $T_\mathrm{settle} = 1$. To this end, the proportional gain of the linear controller is, according to  Equation \refEq{eqn:adrc1_kp}, set to $K_\mathrm{P} = \frac{4}{T_\mathrm{settle}} = 4$. Unless otherwise noted in individual experiments, the observer poles are chosen be $s^\mathrm{ESO}_{1/2} = s^\mathrm{ESO} = 10 \cdot s^\mathrm{CL} = 10 \cdot \left( -K_\mathrm{P} \right) = -40$. The respective values of the observer gains are obtained via Equation \refEq{eqn:adrc1_l_eso}.

Throughout Section \ref{sec:exp_adrc1}, noise-free control variables and ideal measurements will be assumed. Reaction on noisy measurements will be part of further experiments in Section \ref{sec:adrc_discrete_sim}.


\subsubsection{Sensitivity to Process Parameter Variations}
\label{sec:exp_adrc1_kt}

\begin{figure}[t]
 \centering%
 \subfloat[]{%
 \includegraphics[width=0.48\linewidth]{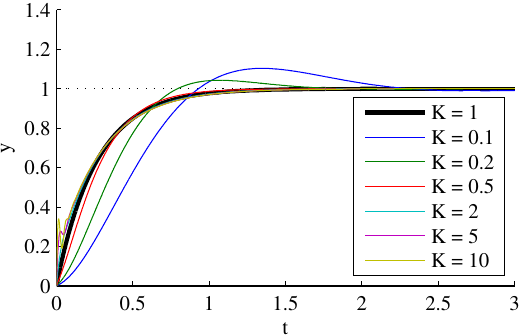}%
 }
 \subfloat[]{%
 \includegraphics[width=0.48\linewidth]{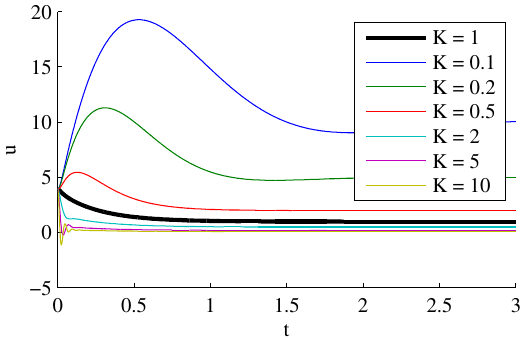}%
 }
 \\
 \subfloat[]{%
 \includegraphics[width=0.48\linewidth]{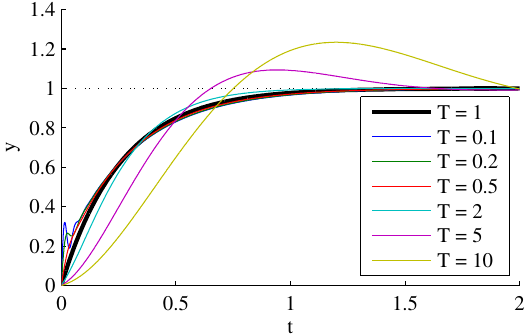}%
 }
 \subfloat[]{%
 \includegraphics[width=0.48\linewidth]{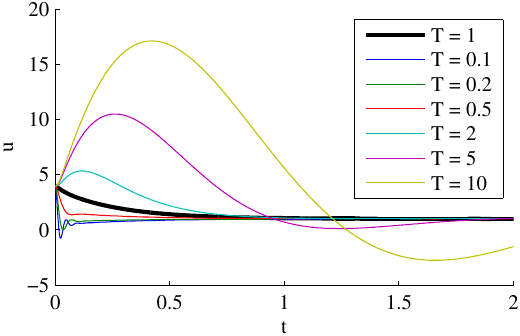}%
 }
 \caption{Experiment~\ref{sec:exp_adrc1_kt}: fixed first-order ADRC controlling first-order process with varying parameters. Nominal process parameters: $K = 1$, $T = 1$. ADRC parameters: $b_0 = \frac{K}{T} = 1$, $T_\mathrm{settle} = 1$, $s^\mathrm{ESO} = 10 \cdot s^\mathrm{CL}$. (\textbf{a}) Variation of $K$, closed loop step response; (\textbf{b})~Controller output $u$ for (\textbf{a}); (\textbf{c}) Variation of $T$, closed loop step response; (\textbf{d}) Controller output $u$ for (\textbf{c}).}
 \label{fig:adrc1_pt1_kt}
\end{figure}

Given the explicit feature of ADRC to cope with modeling errors, our first goal will be to provide visual insights into the control loop behavior under (heavy) variations of process parameters. A series of simulations was run with fixed ADRC parameters as given above and a first-order process with varying parameters, $K$ (DC gain) and $T$ (time constant). In Figure \ref{fig:adrc1_pt1_kt}, the closed loop step responses are displayed on the left- and the according controller outputs on the right-hand side. For both $K$ and $T$, values were chosen from an interval reaching from 10\% of the nominal value to 1000\%, \textit{i.e}., a decrease and increase by factors up to ten.

The results are quite impressive. One can see that the closed loop step response remains similar or nearly identical to the desired behavior (settling time of one second) for most process parameter settings in Figure \ref{fig:adrc1_pt1_kt}. Almost only for the five- and ten-fold increased time constant, larger overshoots become visible. In theory, ideal behavior (\textit{i.e}., almost complete ignorance of parameter variations) can be obtained by placing the observer poles far enough to the left of the closed loop poles, as will be shown in \refSec{sec:exp_adrc1_eso}. Note that we are not constrained by actuator saturation here; this case will be examined in Section \ref{sec:exp_adrc1_kt_ulim}.

\subsubsection{Effect of Observer Pole Locations}
\label{sec:exp_adrc1_eso}

\begin{figure}[t]
 \centering%
 \subfloat[]{%
 \includegraphics[width=0.48\linewidth]{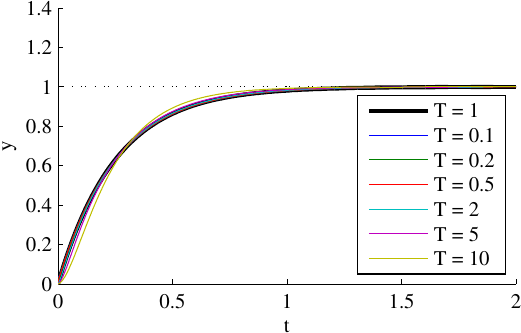}%
 }
 \subfloat[]{%
 \includegraphics[width=0.48\linewidth]{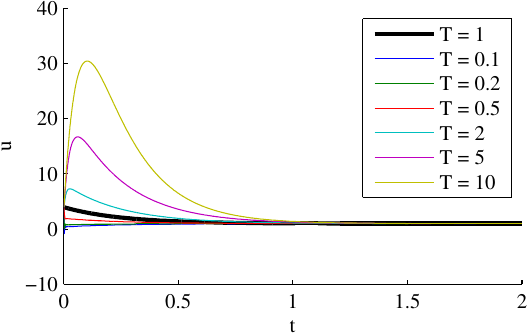}%
 }
 \\
 \subfloat[]{%
 \includegraphics[width=0.48\linewidth]{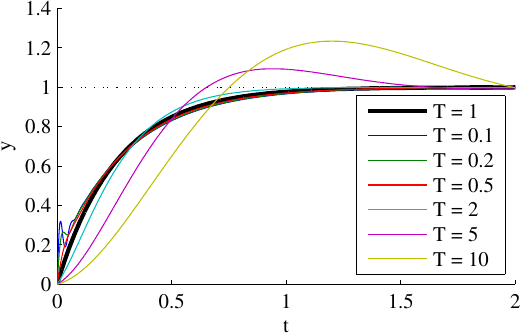}%
 }
 \subfloat[]{%
 \includegraphics[width=0.48\linewidth]{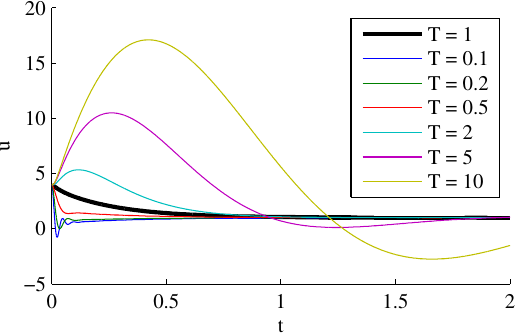}%
 }
 \\
 \subfloat[]{%
 \includegraphics[width=0.48\linewidth]{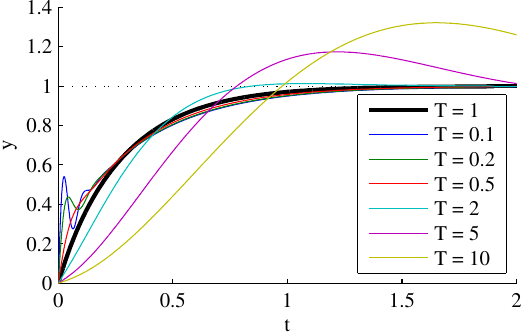}%
 }
 \subfloat[]{%
 \includegraphics[width=0.48\linewidth]{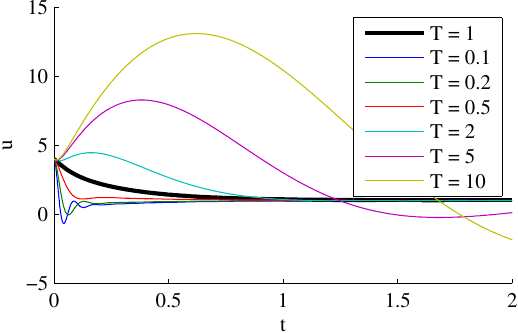}%
 }
 \caption{Experiment~\ref{sec:exp_adrc1_eso}: effect of pole locations for the extended state observer (ESO) on a fixed first-order ADRC controlling first-order process with varying parameter, $T$. Nominal process parameters: $K = 1$, $T = 1$. ADRC parameters: $b_0 = \frac{K}{T} = 1$, $T_\mathrm{settle} = 1$, $s^\mathrm{ESO}$ varying. (\textbf{a}) Closed loop step response, $s^\mathrm{ESO} = 100 \cdot s^\mathrm{CL}$; (\textbf{b}) Controller output $u$ for (\textbf{a}); (\textbf{c}) Closed loop step response, $s^\mathrm{ESO} = 10 \cdot s^\mathrm{CL}$; (\textbf{d}) Controller output $u$ for (\textbf{c}); (\textbf{e}) Closed loop step response, $s^\mathrm{ESO} = 5 \cdot s^\mathrm{CL}$; (\textbf{f}) Controller output $u$ for (\textbf{e}).}
 \label{fig:adrc1_pt1_t_eso}
\end{figure}

In the previous Section~\ref{sec:exp_adrc1_kt}, the observer poles were placed ten times faster than the closed loop pole ($s^\mathrm{ESO}_{1/2} = s^\mathrm{ESO} = 10 \cdot s^\mathrm{CL}$). How does the choice of this factor influence the behavior and abilities regarding process parameter variations? We will repeat the simulations with varying time constant, $T$, of process both with slower and faster observers. To demonstrate a rather extreme case, as well, we will, for the faster setting, apply a factor of $s^\mathrm{ESO} = 100 \cdot s^\mathrm{CL}$.

The results in Figure \ref{fig:adrc1_pt1_t_eso} are ordered from fastest (top) to slowest observer setting (bottom). For the fastest setting, the theoretical ability to almost completely ignore any modeling error is confirmed. In order to achieve this behavior in practice, the actuator must be fast enough and must not saturate within the desired range of parameter variations. For slower observer settings, the actual closed loop dynamics increasingly differ from the desired dynamics, noticeable especially from larger overshoots for process variations with stronger low-pass character.


\subsubsection{Effect of Actuator Saturation}
\label{sec:exp_adrc1_kt_ulim}

\begin{figure}[t!]
 \centering%
 \includegraphics{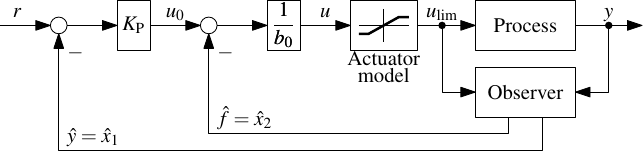}%
 \caption{Control loop structure of first-order ADRC considering actuator saturation.}
 \label{fig:adrc1_structure_ulim}
\end{figure}

\begin{figure}[t!]
 \centering%
 \subfloat[]{%
 \includegraphics[width=0.48\linewidth]{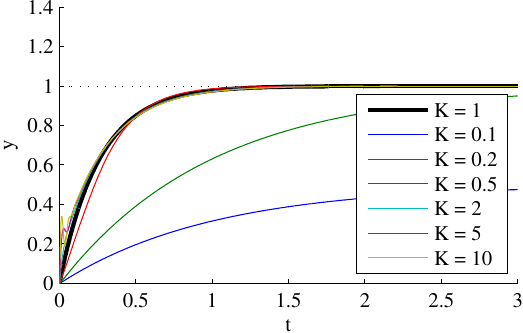}%
 }
 \subfloat[]{%
 \includegraphics[width=0.48\linewidth]{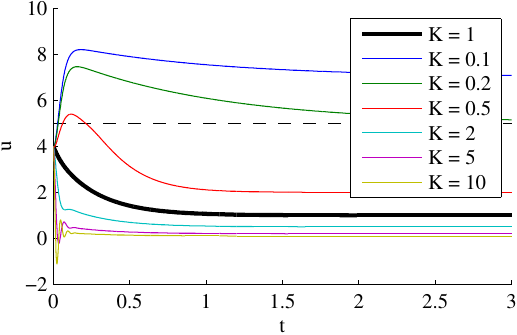}%
 }
 \\
 \subfloat[]{%
 \includegraphics[width=0.48\linewidth]{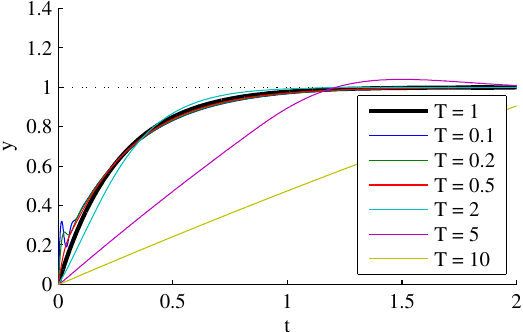}%
 }
 \subfloat[]{%
 \includegraphics[width=0.48\linewidth]{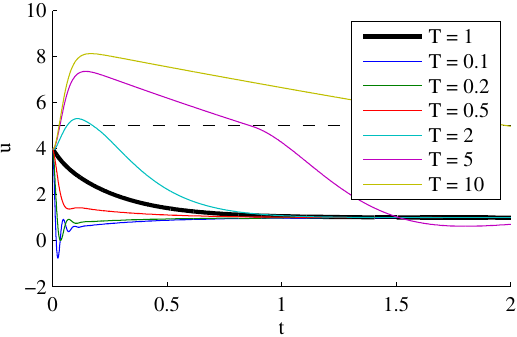}%
 }
 \caption{Experiment~\ref{sec:exp_adrc1_kt_ulim}: effect of actuator saturation, $|u_\mathrm{lim}| \le 5$, on fixed a first-order ADRC controlling first-order process with varying parameters. Nominal process parameters: $K = 1$, $T = 1$. ADRC parameters: $b_0 = \frac{K}{T} = 1$, $T_\mathrm{settle} = 1$, $s^\mathrm{ESO} = 10 \cdot s^\mathrm{CL}$. On the right-hand side, the controller outputs are shown before the limitation (dashed line) takes effect. (\textbf{a}) Variation of $K$, closed loop step response; (\textbf{b}) Controller output $u$ for (\textbf{a}); (\textbf{c}) Variation of $T$, closed loop step response; (\textbf{d}) Controller output $u$ for (\textbf{c}).}
 \label{fig:adrc1_pt1_kt_ulim}
\end{figure}

In practice, the abilities of any controller are tied to limitations of the actuator, \textit{i.e}., its dynamics and the realizable range of values of the actuating variable. While the actuator dynamics can be viewed as part of the process dynamics during controller design, one has to take possible effects of actuator saturation into account.

If parameters of our example process change, the control loop behavior may be influenced or limited by actuator saturation in different ways: If the process becomes slower (\textit{i.e}., the time constant $T$ increases), actuator saturation will increase the settling time. If, on the other hand, the DC gain of the process decreases, the control loop may be not be able to reach the reference value under actuator saturation.

In classical PID-type control, some sort of anti-windup strategy would be required to prevent the side-effects of actuator saturation. We will see in the experiments of this section that for ADRC, those effects can be overcome very simply by feeding the state observer with the limited actuating variable, $u_\mathrm{lim}(t)$, instead of $u(t)$, either by a measured value or by an actuator model, \textit{cf}.\ Figure \ref{fig:adrc1_structure_ulim}.

To demonstrate the behavior of ADRC under actuator saturation, the experiments of Section \ref{sec:exp_adrc1_kt} will now be repeated under an (arbitrarily chosen) limitation of the actuating variable, $|u_\mathrm{lim}| \le 5$. \refFigBegin{fig:adrc1_pt1_kt_ulim} shows the control loop behavior with varying process parameters, $K$ and $T$. On the right-hand side of Figure \ref{fig:adrc1_pt1_kt_ulim}, the controller outputs are shown before being fed into the actuator model, \textit{i.e}., before the limitation becomes effective. One can see that for reduced process gains, $K \le 0.2$, the reference value cannot be reached anymore. From the respective controller outputs, it becomes apparent that $u(t)$ does not wind up when actuator saturation takes effect, but converges to a steady-state value.

For slower process dynamics, $T = 5$ and $T = 10$, the settling time increases considerably, yet there is almost no overshoot visible. Since the actuator is saturated, this already is the fastest possible step response. Obviously, the controller recovers very well from periods of actuator saturation. To summarize, for practical implementations of ADRC, this means that apart from the small modification shown in Figure \ref{fig:adrc1_structure_ulim}, there are no further anti-windup measures necessary.


\begin{figure}[t]
 \centering%
 \subfloat[]{%
 \includegraphics[width=0.48\linewidth]{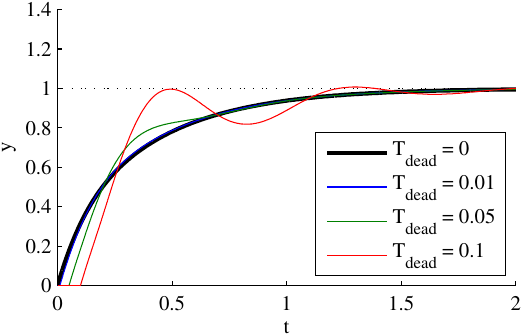}%
 }
 \subfloat[]{%
 \includegraphics[width=0.48\linewidth]{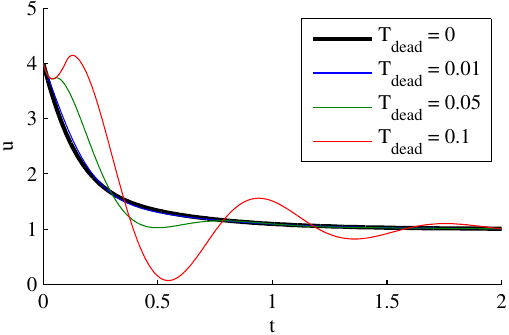}%
 }
 \\
 \subfloat[]{%
 \includegraphics[width=0.48\linewidth]{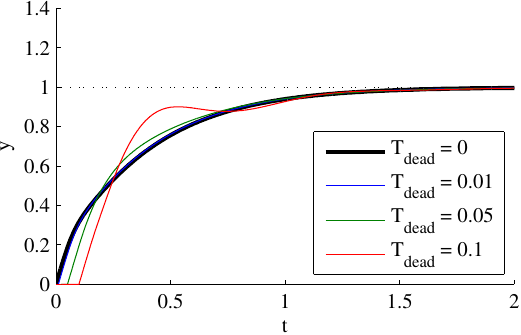}%
 \label{fig:adrc1_pt1_dt_1y}%
 }
 \subfloat[]{%
 \includegraphics[width=0.48\linewidth]{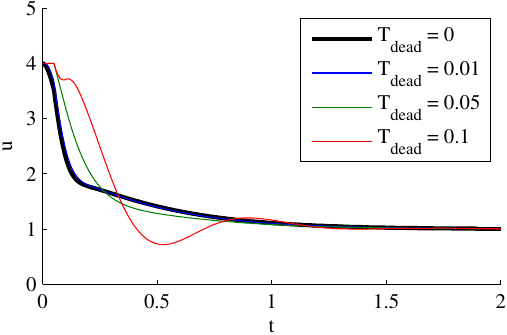}%
 \label{fig:adrc1_pt1_dt_1u}%
 }
 \caption{Experiment~\ref{sec:exp_adrc1_dt}: effect of dead time, $T_\mathrm{dead}$, on a first-order ADRC controlling a process with first-order dynamics ($K = 1$, $T = 1$) and unknown dead time. ADRC parameters: $b_0 = \frac{K}{T} = 1$, $T_\mathrm{settle} = 1$, $s^\mathrm{ESO} = 2 \cdot s^\mathrm{CL}$. In (\textbf{c}) and (\textbf{d}), a fixed dead time, $T^\mathrm{ESO}_\mathrm{dead}$, was incorporated into the observer to improve the controller behavior. (\textbf{a}) Variation of $T_\mathrm{dead}$, closed loop step response; (\textbf{b}) Controller output $u$ for (\textbf{a}); (\textbf{c}) Closed loop step response, observer with $T^\mathrm{ESO}_\mathrm{dead} = 0.05$; (\textbf{d}) Controller output $u$ for (\textbf{c}).
 }
 \label{fig:adrc1_pt1_dt}
\end{figure}

\subsubsection{Effect of Dead Time}
\label{sec:exp_adrc1_dt}

Many practical processes with dominating first-order behavior do exhibit a dead time. While there are many specialized model-based approaches to control such processes, we are interested in how ADRC will handle an unknown---albeit small---amount of dead time in the control loop.

In Figure \ref{fig:adrc1_pt1_dt}, simulations were performed as in previous experiments, and a dead time, $T_\mathrm{dead}$, was added to the process with $T_\mathrm{dead} \le 0.1$, \textit{i.e}., up to 10\,\%, compared to the time constant, $T$, of the process. As expected, oscillations are inevitably starting to appear with increasing $T_\mathrm{dead}$, especially in the controller output. However, this situation can be improved if the dead time of the process is---at least approximately---known. An easy way of incorporating small dead times into ADRC can be found by delaying the controller output fed into the observer by $T^\mathrm{ESO}_\mathrm{dead}$, \textit{i.e}., using $u(t-T^\mathrm{ESO}_\mathrm{dead})$ instead of $u(t)$ in Equation \refEq{eqn:pt1_eso}. In Figure \ref{fig:adrc1_pt1_dt_1y} and \subref{fig:adrc1_pt1_dt_1u}, this approach was implemented using $T^\mathrm{ESO}_\mathrm{dead} = 0.05$. Clearly, the oscillations in the controller output are less prominent, even if $T^\mathrm{ESO}_\mathrm{dead}$ does not match the actual dead time.


\subsubsection{Effect of Structural Uncertainties}
\label{sec:exp_adrc1_vs}

\begin{figure}[t]
 \centering%
 \subfloat[]{%
 \includegraphics[width=0.48\linewidth]{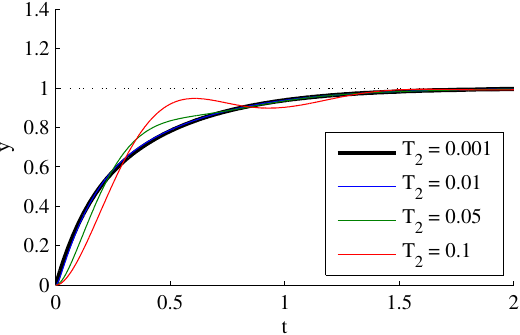}%
 }
 \subfloat[]{%
 \includegraphics[width=0.48\linewidth]{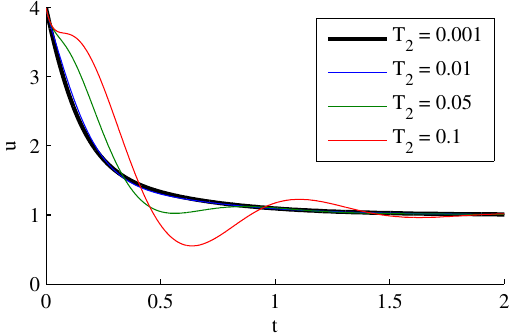}%
 }
 \caption{Experiment~\ref{sec:exp_adrc1_vs}: effect of structural uncertainties on a first-order ADRC controlling a process with dominating first-order behavior ($K = 1$, $T = 1$) and higher-order dynamics caused by an unknown second pole at $s = -1/T_2$. ADRC parameters: $b_0 = \frac{K}{T} = 1$, $T_\mathrm{settle} = 1$, $s^\mathrm{ESO} = 2 \cdot s^\mathrm{CL}$. (\textbf{a}) Variation of $T_2$, closed loop step response; (\textbf{b}) Controller output $u$ for (\textbf{a}).}
 \label{fig:adrc1_pt1_vs}
\end{figure}

In the experiments carried out so far in this section, it was assumed that our process could be reasonably well described by a first-order model. In practice, such a first-order model almost always results from neglecting higher-order dynamics, e.g.\ of the actuator. While it could already be seen that ADRC can handle variations of parameters very well, how does it behave if higher-order dynamics become unexpectedly visible in the process?

To demonstrate this behavior, a second pole was added to the process in the simulations from Figure~\ref{fig:adrc1_pt1_vs}, resulting in a second time constant, $T_2 \le 0.1$, \textit{i.e}., up to 10\% of the dominant time constant, $T$. One can see that some oscillations start to appear during the transient as the higher-order pole approaches the dominant pole. While these results are acceptable, further simulations showed that ADRC did not provide an advantage comparing to standard PI controllers as large as it does in the case of parameter robustness, \textit{cf}.\ Section \ref{sec:exp_adrc1_pi}.


\subsubsection{Comparison to PI Control}
\label{sec:exp_adrc1_pi}

\begin{figure}[t]
 \centering%
 \subfloat[]{%
 \includegraphics[width=0.48\linewidth]{adrc1_pt1_K_y.pdf}%
 }
 \subfloat[]{%
 \includegraphics[width=0.48\linewidth]{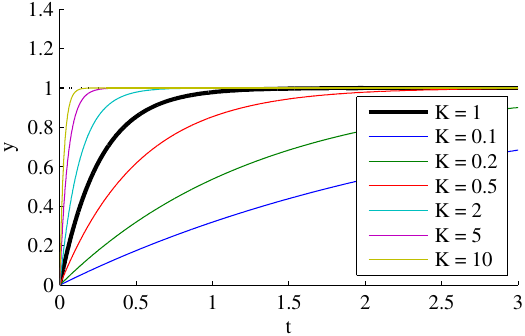}%
 }
 \\
 \subfloat[]{%
 \includegraphics[width=0.48\linewidth]{adrc1_pt1_T_y.pdf}%
 }
 \subfloat[]{%
 \includegraphics[width=0.48\linewidth]{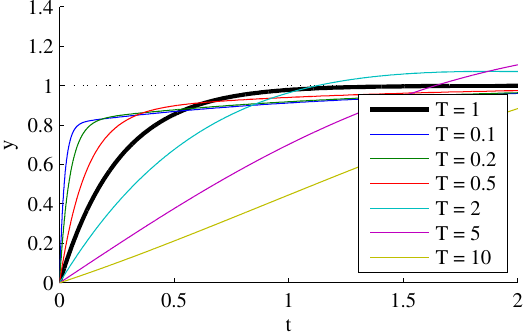}%
 }
 \caption{Experiment~\ref{sec:exp_adrc1_pi}: comparison of first-order ADRC and PI controller faced with varying process parameters, $K$ and $T$. Nominal process and ADRC parameters are as throughout Section \ref{sec:exp_adrc1}. For each combination, the closed loop step response is shown. (\textbf{a}) Variation of $K$, ADRC; (\textbf{b}) Variation of $K$, PI; (\textbf{c}) Variation of $T$, ADRC; (\textbf{d}) Variation of $T$, PI.}
 \label{fig:adrc1_pt1_pi}
\end{figure}

Given the ubiquity of PID-type controllers in industrial practice, how does the standard approach keep up against ADRC? To that end, we will repeat the experiment regarding sensitivity towards process parameter variations and compare the ADRC results to a PI controller.

For the first-order process, a PI controller is sufficient to achieve any desired first-order closed loop behavior. In order to obtain comparable results, the PI controller is designed for nearly identical closed loop dynamics as the ADRC by aiming for the same settling time and placing a zero on the pole of the first-order process:
\begin{equation*}
C_\mathrm{PI}(s) = K_\mathrm{P} + \D\frac{K_\mathrm{I}}{s}
\quad \text{with} \quad
K_\mathrm{I} = \frac{3.85}{K \cdot T_\mathrm{settle}} = 3.85
\quad \text{and} \quad
K_\mathrm{P} = K_\mathrm{I} \cdot T = 3.85
\end{equation*}

The simulation results in Figure \ref{fig:adrc1_pt1_pi} clearly demonstrate the ability of the ADRC approach to keep the closed loop dynamics similar, even under major parameter variations. The PI controller delivers dynamics that vary heavily, as do the process parameters.


\subsubsection{Disturbance Rejection of ADRC and PI}
\label{sec:exp_adrc1_disturbance}

As a final experiment for the continuous-time first-order ADRC, we want to examine the disturbance rejection abilities by injecting an input disturbance into the process for both ADRC and the PI controller from Section \ref{sec:exp_adrc1_pi}.

\begin{figure}[t]
 \centering%
 \subfloat[]{%
 \includegraphics[width=0.48\linewidth]{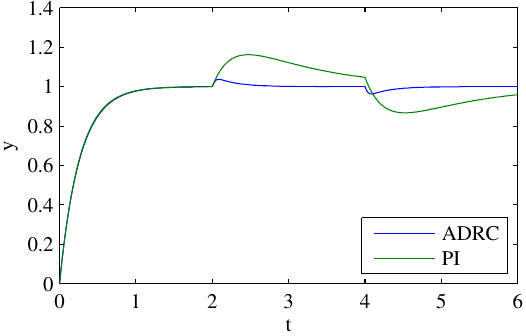}%
 }
 \subfloat[]{%
 \includegraphics[width=0.48\linewidth]{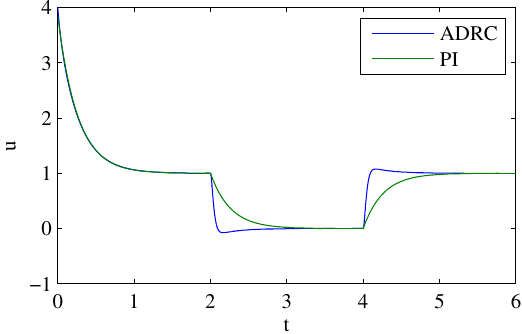}%
 }
 \caption{Experiment~\ref{sec:exp_adrc1_disturbance}: comparison of disturbance rejection behavior (first-order ADRC \textit{vs}.\ PI controller). Process and ADRC parameters are as throughout Section \ref{sec:exp_adrc1}. Input disturbance, $d = 1$, was effective from $t = 2$ until $t = 4$. (\textbf{a}) Step response and reaction on disturbance; (\textbf{b}) Controller output $u$ for (\textbf{a}).}
 \label{fig:adrc1_pt1_z}
\end{figure}

On the left-hand side of Figure \ref{fig:adrc1_pt1_z}, a closed loop step response is shown. The input disturbance is effective during the period from $t = 2$ until $t = 4$. While both controllers were tuned for the same reaction on setpoint changes, the impact of the disturbance is compensated for much faster by ADRC compared to the PI controller. In classical control, one would need to tune the PI controller much more aggressively and add a setpoint filter or follow another 2DOF approach in order to obtain similar results \cite{Araki:2003}.

\subsection{Second-Order ADRC with a Second-Order Process}
\label{sec:exp_adrc2}

The second-order ADRC will be examined using a second-order process with the following nominal~parameters:
\begin{equation*}
P(s) = \D\frac{y(s)}{u(s)} = \D\frac{K}{T^2 s + 2DTs + 1}
\quad \text{with} \quad
K = 1, \quad D = 1 \quad \text{and} \quad T = 1
\end{equation*}

Again, perfect knowledge of the process is assumed, $b_0 = \frac{K}{T^2} = 1$, but then $b_0$ is left unchanged throughout Section \ref{sec:exp_adrc2}. The desired closed loop settling time is $T_\mathrm{settle} = 5$. The parameters of the PD controller are, following Equation \refEq{eqn:adrc2_kp_kd}, set to $K_\mathrm{P} = \left(\frac{6}{T_\mathrm{settle}}\right)^2 = 1.44$ and $K_\mathrm{D} = \frac{12}{T_\mathrm{settle}} = 2.4$. Unless otherwise noted in individual experiments, the observer poles are chosen to be $s^\mathrm{ESO}_{1/2/3} = s^\mathrm{ESO} = 10 \cdot s^\mathrm{CL} = 10 \cdot \left( -\frac{6}{T_\mathrm{settle}} \right) = -12$. The respective values of the observer gains, $l_{1/2/3}$, are obtained via Equation \refEq{eqn:adrc2_l_eso}. As in Section \ref{sec:exp_adrc1}, throughout this section, noise-free control variables and ideal measurements will be~assumed.


\begin{figure}[t]
 \centering%
 \subfloat[]{%
 \includegraphics[width=0.48\linewidth]{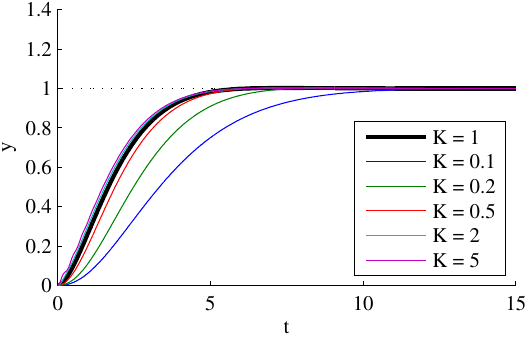}%
 }
 \subfloat[]{%
 \includegraphics[width=0.48\linewidth]{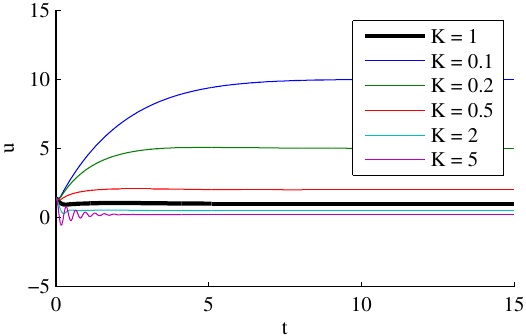}%
 }
 \\
 \subfloat[]{%
 \includegraphics[width=0.48\linewidth]{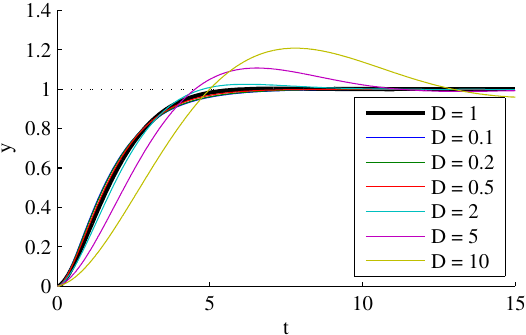}%
 }
 \subfloat[]{%
 \includegraphics[width=0.48\linewidth]{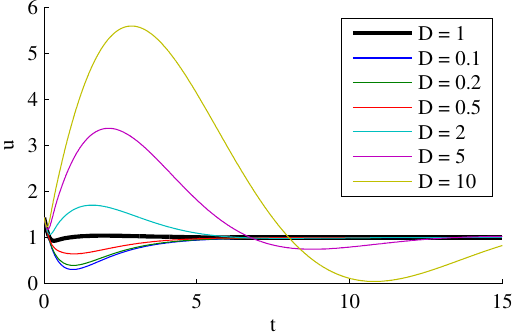}%
 }
 \\
 \subfloat[]{%
 \includegraphics[width=0.48\linewidth]{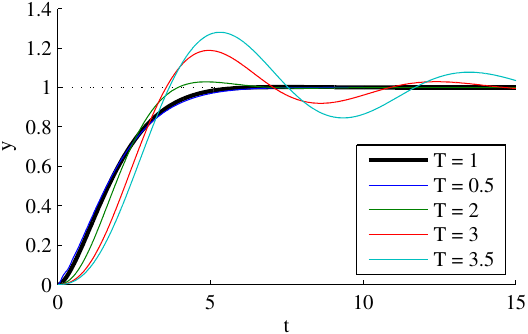}%
 }
 \subfloat[]{%
 \includegraphics[width=0.48\linewidth]{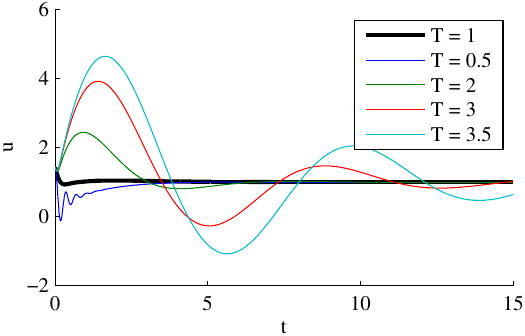}%
 }
 \caption{Experiment~\ref{sec:exp_adrc2_kdt}: a fixed second-order ADRC controlling second-order process with varying parameters. Nominal process parameters: $K = 1$, $D = 1$, $T = 1$. ADRC parameters: $b_0 = \frac{K}{T^2} = 1$, $T_\mathrm{settle} = 5$, $s^\mathrm{ESO} = 10 \cdot s^\mathrm{CL}$. (\textbf{a}) Variation of $K$, closed loop step response; (\textbf{b}) Controller output $u$ for (\textbf{a}); (\textbf{c}) Variation of $D$, closed loop step response; (\textbf{d}) Controller output $u$ for (\textbf{c}); (\textbf{e}) Variation of $T$, closed loop step response; (\textbf{f}) Controller output $u$ for (\textbf{e}).}
 \label{fig:adrc2_pt2_kdt}
\end{figure}

\subsubsection{Sensitivity to Process Parameter Variations}
\label{sec:exp_adrc2_kdt}

For the second-order ADRC, sensitivity to variations of the process parameters will be examined firstly. Deviations from the original DC gain ($K = 1$) were made in a range of $10\%$ to $500\%$ by setting $K$ to one of the values $0.1$, $0.2$, $0.5$, $2$ and $5$. The damping factor, $D$, of the process was varied within a range of $10\%$ to $1000\%$ of the original value, $D = 1$, using settings $0.1$, $0.2$, $0.5$, $2$, $5$ and $10$. Finally, the time constant, $T$, of the second-order process was varied in a range of $50\%$ to $350\%$ of the original value, $T = 1$, using the values $0.5$, $2$, $3$ and $3.5$.

From the results in Figure \ref{fig:adrc2_pt2_kdt}, one can see that the closed loop behavior is almost not at all affected, even by relatively large changes of $K$ and $D$. Only for very small values of $K$, the step response becomes slower; and for very large values of the damping, $D$, some overshoot becomes visible. Changes in the time constant, $T$, do, for larger values, increase overshoot and oscillations more than in the first-order case.

In order to fully appreciate the robustness towards parameter changes of the process, one has to compare these results against standard controllers, such as PID, as will be done in Section \ref{sec:exp_adrc2_pid}.


\begin{figure}[t]
 \centering%
 \subfloat[]{%
 \includegraphics[width=0.48\linewidth]{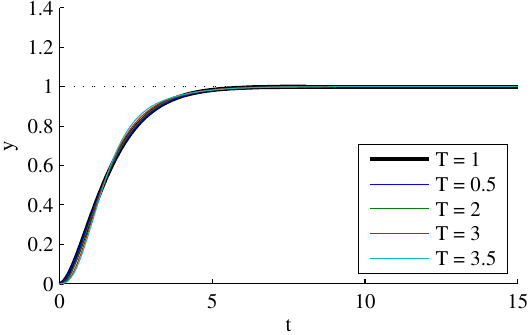}%
 }
 \subfloat[]{%
 \includegraphics[width=0.48\linewidth]{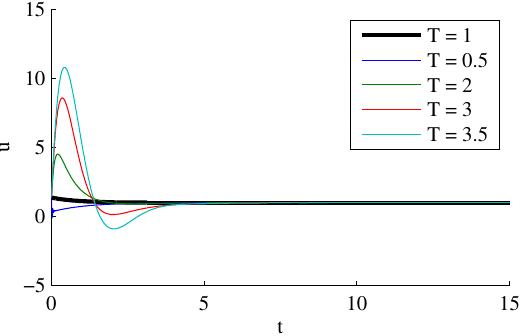}%
 }
 \\
 \subfloat[]{%
 \includegraphics[width=0.48\linewidth]{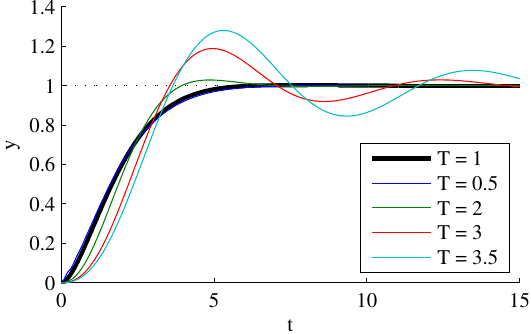}%
 }
 \subfloat[]{%
 \includegraphics[width=0.48\linewidth]{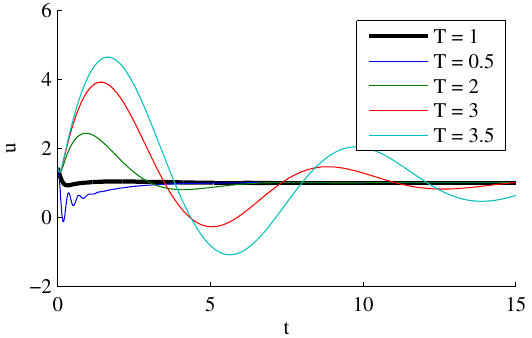}%
 }
 \\
 \subfloat[]{%
 \includegraphics[width=0.48\linewidth]{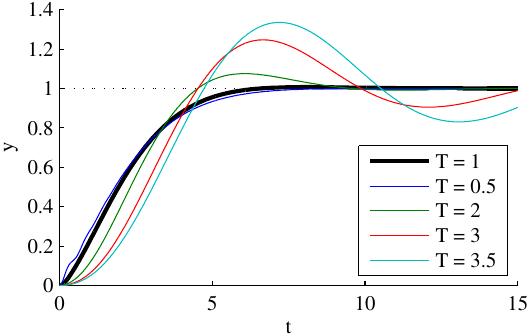}%
 }
 \subfloat[]{%
 \includegraphics[width=0.48\linewidth]{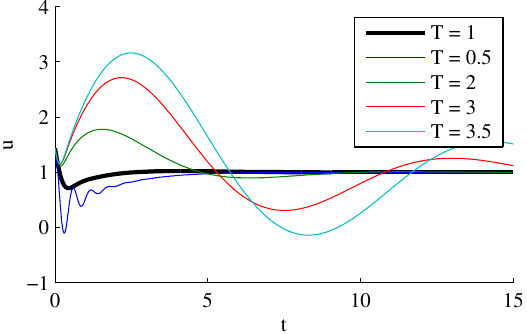}%
 }
 \caption{Experiment~\ref{sec:exp_adrc2_eso}: effect of observer pole locations for a second-order ADRC controlling second-order processes with varying parameter, $T$. Nominal process parameters: $K = 1$, $D = 1$, $T = 1$. ADRC parameters: $b_0 = \frac{K}{T^2} = 1$, $T_\mathrm{settle} = 5$, $s^\mathrm{ESO}$ varying. (\textbf{a}) Closed loop step response, $s^\mathrm{ESO} = 100 \cdot s^\mathrm{CL}$; (\textbf{b}) Controller output $u$ for (\textbf{a}); (\textbf{c}) Closed loop step response, $s^\mathrm{ESO} = 10 \cdot s^\mathrm{CL}$; (\textbf{d}) Controller output $u$ for (\textbf{c}); (\textbf{e}) Closed loop step response, $s^\mathrm{ESO} = 5 \cdot s^\mathrm{CL}$; (\textbf{f}) Controller output $u$ for (\textbf{e}).}
 \label{fig:adrc2_pt2_t_eso}
\end{figure}

\subsubsection{Effect of Observer Pole Locations}
\label{sec:exp_adrc2_eso}

In the previous section, we saw that changes in the process time constant, $T$, affected the closed loop behavior stronger than changes in $K$ and $D$. We will therefore demonstrate the influence of the observer poles on the sensitivity by repeating the experiments with varying $T$ for different observer pole locations.

In the previous experiments, the observer poles were set ten times to the left of the closed loop poles in the $s$-plane, \textit{i.e}., $s^\mathrm{ESO} = 10 \cdot s^\mathrm{CL}$. Here, both slower and faster observers will be examined, as well, by setting $s^\mathrm{ESO} = 100 \cdot s^\mathrm{CL}$ and $s^\mathrm{ESO} = 5 \cdot s^\mathrm{CL}$.

As visible in the simulation results in Figure \ref{fig:adrc2_pt2_t_eso}, almost ideal behavior can---at least in theory---be obtained by placing the observer poles far enough left in the $s$-plane, \textit{i.e}., choosing large factors $k$ in $s^\mathrm{ESO} = k \cdot s^\mathrm{CL}$. Of course, this comes at the price of the need for faster actuators and larger controller outputs. Furthermore---as we will see in Section \ref{sec:adrc_discrete_sim}---faster observers are more sensitive to measurement noise and will be limited by sample time restrictions in discrete time implementations.


\begin{figure}[t]
 \centering%
 \subfloat[]{%
 \includegraphics[width=0.48\linewidth]{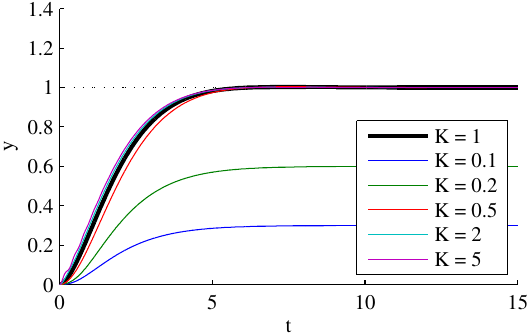}%
 }
 \subfloat[]{%
 \includegraphics[width=0.48\linewidth]{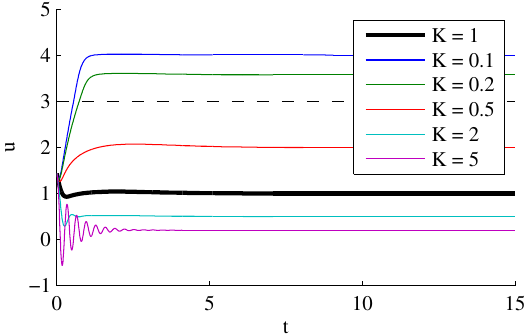}%
 \label{fig:adrc2_pt2_kdt_ulim_b}%
 }
 \\
 \subfloat[]{%
 \includegraphics[width=0.48\linewidth]{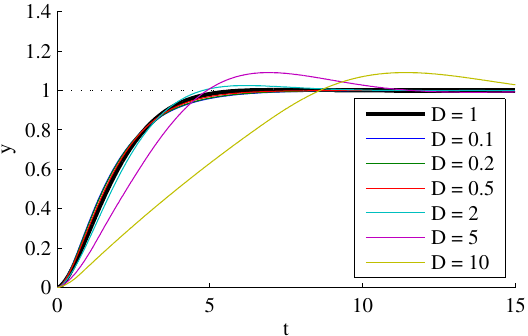}%
 }
 \subfloat[]{%
 \includegraphics[width=0.48\linewidth]{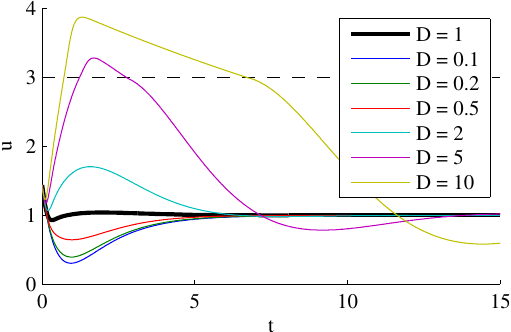}%
 }
 \\
 \subfloat[]{%
 \includegraphics[width=0.48\linewidth]{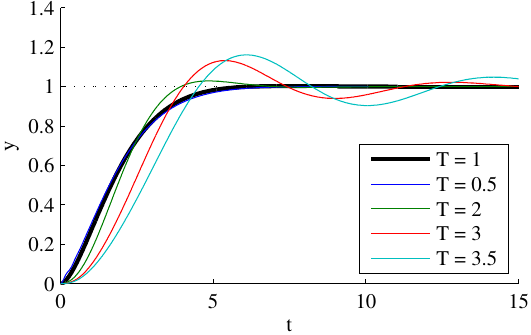}%
 }
 \subfloat[]{%
 \includegraphics[width=0.48\linewidth]{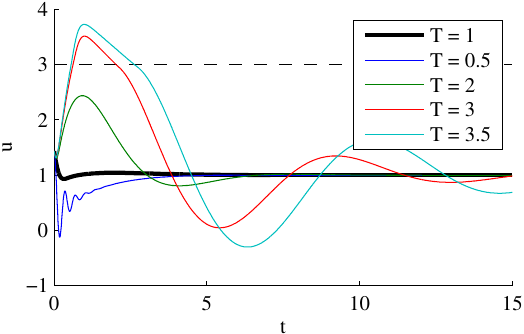}%
 }
 \caption{Experiment~\ref{sec:exp_adrc2_kdt_ulim}: effect of actuator saturation, $|u_\mathrm{lim}| \le 3$, on a fixed second-order ADRC controlling a second-order process with varying parameters. Nominal process parameters: $K = 1$, $D = 1$, $T = 1$. ADRC parameters: $b_0 = \frac{K}{T^2} = 1$, $T_\mathrm{settle} = 5$, $s^\mathrm{ESO} = 10 \cdot s^\mathrm{CL}$. The controller outputs are shown before the limitation (dashed line) takes effect. (\textbf{a}) Variation of $K$, closed loop step response; (\textbf{b}) Controller output $u$ for (\textbf{a}); (\textbf{c}) Variation of $D$, closed loop step response; (\textbf{d}) Controller output $u$ for (\textbf{c}); (\textbf{e}) Variation of $T$, closed loop step response; (\textbf{f}) Controller output $u$ for (\textbf{e}).}
 \label{fig:adrc2_pt2_kdt_ulim}
\end{figure}

\subsubsection{Effect of Actuator Saturation}
\label{sec:exp_adrc2_kdt_ulim}

Analogously to Section \ref{sec:exp_adrc1_kt_ulim}, the experiments were carried out by extending the controller structure, such that the saturated controller output, $u_\mathrm{lim}(t)$, was fed back to the observer instead of $u(t)$ (compare Figure \ref{fig:adrc1_structure_ulim} for the first-order case). The resulting step responses and controller outputs for varying process parameters can be found in Figure \ref{fig:adrc2_pt2_kdt_ulim}. As expected, lowering the DC gain, $K$, of the process forced the controller into saturation, such that the desired process output could not be reached anymore. It has to be stressed again that the controller output did not wind up, as visible in Figure \ref{fig:adrc2_pt2_kdt_ulim}.

When decelerating the process (increased damping $D$), the closed loop dynamics suffered when the controller output ran into saturation (increased settling time), but there were no additional oscillations after recovering from saturation, as known from classical PID control without anti-windup measures.


\subsubsection{Effect of Dead Time}
\label{sec:exp_adrc2_dt}

\begin{figure}[t]
 \centering%
 \subfloat[]{%
 \includegraphics[width=0.48\linewidth]{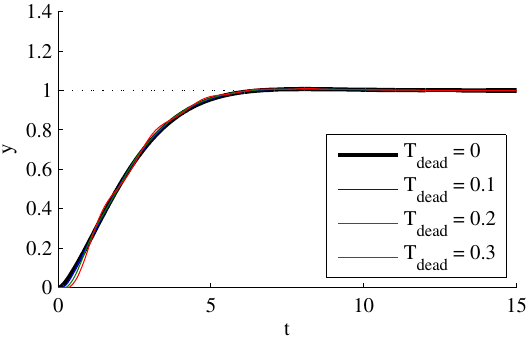}%
 \label{fig:adrc2_pt2_dt_y}%
 }
 \subfloat[]{%
 \includegraphics[width=0.48\linewidth]{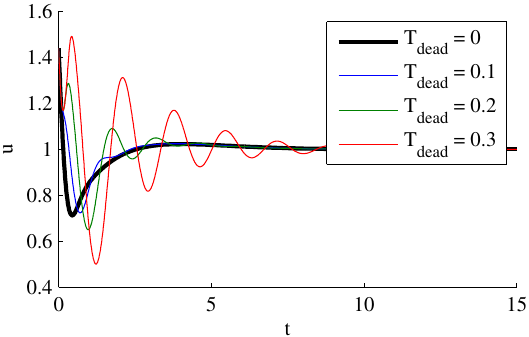}%
 \label{fig:adrc2_pt2_dt_u}%
 }
 \\
 \subfloat[]{%
 \includegraphics[width=0.48\linewidth]{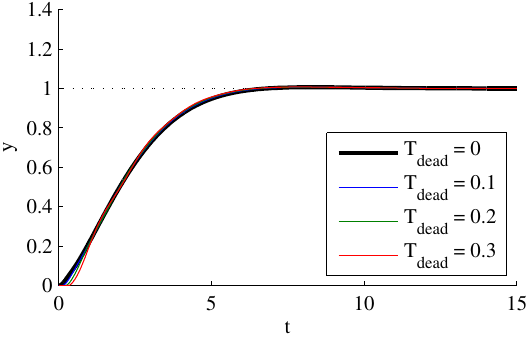}%
 \label{fig:adrc2_pt2_dt_y1}%
 }
 \subfloat[]{%
 \includegraphics[width=0.48\linewidth]{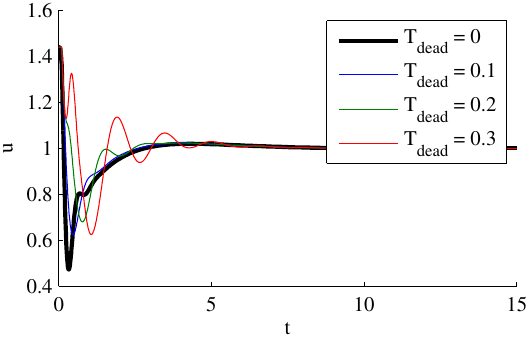}%
 \label{fig:adrc2_pt2_dt_u1}%
 }
 \caption{Experiment~\ref{sec:exp_adrc2_dt}: effect of dead time, $T_\mathrm{dead}$, on a second-order ADRC controlling a process with second-order dynamics ($K = 1$, $D = 1$, $T = 1$) and unknown dead time. ADRC parameters: $b_0 = \frac{K}{T^2} = 1$, $T_\mathrm{settle} = 5$, $s^\mathrm{ESO} = 5 \cdot s^\mathrm{CL}$. In (\textbf{c}) and (\textbf{d}), a fixed dead time, $T^\mathrm{ESO}_\mathrm{dead}$, was incorporated into the observer to improve the controller behavior. (\textbf{a}) Variation of $T_\mathrm{dead}$, closed loop step response; (\textbf{b}) Controller output $u$ for (\textbf{a}); (\textbf{c}) Closed loop step response, observer with $T^\mathrm{ESO}_\mathrm{dead} = 0.1$; (\textbf{d}) Controller output $u$ for (\textbf{a}).
 }
 \label{fig:adrc2_pt2_dt}
\end{figure}

Similar to Section \ref{sec:exp_adrc1_dt}, the second-order ADRC was confronted with an unknown dead time in the process with $T_\mathrm{dead} \le 0.3$. One can see from Figure \ref{fig:adrc2_pt2_dt_y} that the controller and observer work very well together, such that the output is hardly being affected by the dead time, which is also an improvement compared to the first-order case in Section \ref{sec:exp_adrc1_dt}. In the controller output in Figure \ref{fig:adrc2_pt2_dt_u}, however, oscillations become increasingly visible with larger values of $T_\mathrm{dead}$. Again, this situation can be improved by delaying the input of the observer, \textit{i.e}., using $u(t-T_\mathrm{dead}^\mathrm{ESO})$ instead of $u(t)$ in Equation~\refEq{eqn:pt2_eso}. For a fixed value, $T_\mathrm{dead}^\mathrm{ESO} = 0.1$, the controller behavior is shown in Figure \ref{fig:adrc2_pt2_dt_y1} and \subref{fig:adrc2_pt2_dt_u1}, where the controller oscillations are already significantly reduced, even if $T_\mathrm{dead}^\mathrm{ESO}$ does not match the actual dead time.


\subsubsection{Effect of Structural Uncertainties}
\label{sec:exp_adrc2_vs}

\begin{figure}[t]
 \centering%
 \subfloat[]{%
 \includegraphics[width=0.48\linewidth]{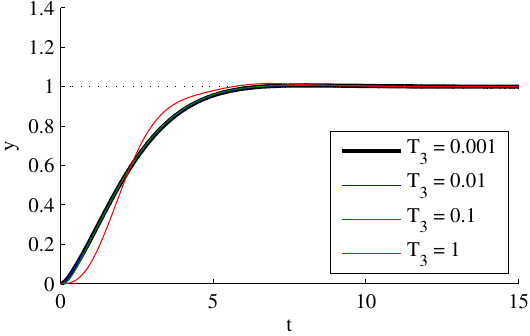}%
 }
 \subfloat[]{%
 \includegraphics[width=0.48\linewidth]{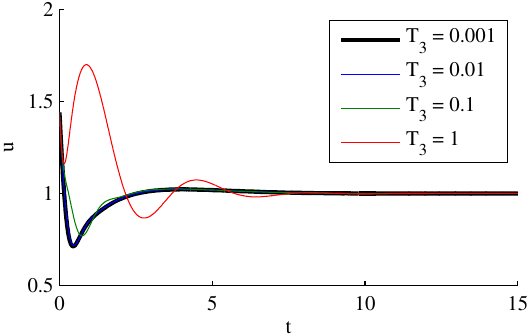}%
 }
 \caption{Experiment~\ref{sec:exp_adrc2_vs}: effect of structural uncertainties on a second-order ADRC controlling a process with dominating second-order behavior ($K = 1$, $D = 1$, $T = 1$) and higher-order dynamics caused by an unknown third pole at $s = -1/T_3$. ADRC parameters: $b_0 = \frac{K}{T^2} = 1$, $T_\mathrm{settle} = 5$, $s^\mathrm{ESO} = 5 \cdot s^\mathrm{CL}$. (\textbf{a}) Variation of $T_3$, closed loop step response; (\textbf{b}) Controller output $u$ for (\textbf{a}).}
 \label{fig:adrc2_pt2_vs}
\end{figure}

Besides unknown dead times, the controller may be faced with higher-order dynamics in the process. Therefore, the effect of an unknown third pole in the process was examined in Figure \ref{fig:adrc2_pt2_vs}. The time constant, $T_3$, of the third pole was varied within $0.001 \le T_3 \le 1$, \textit{i.e}., the third pole was identical to the two known poles of the plant in the extreme case. In contrast to the first-order ADRC in Section~\ref{sec:exp_adrc1_vs}, the second-order ADRC proved to have very good robustness against an unknown higher-order dynamics, even in the more challenging cases of Figure \ref{fig:adrc2_pt2_vs}.

\subsubsection{Comparison to PI and PID Control}
\label{sec:exp_adrc2_pid}

To view and assess the abilities of the second-order ADRC from a perspective of classical control, we will now employ standard PI and PID controllers and---after fixing their parameters---expose them to the same process parameter variations, as done in Section \ref{sec:exp_adrc2_kdt}. To ensure comparability, all controllers are being designed for the same closed loop dynamics (settling time $T_\mathrm{settle} = 5$, no overshoot) using the nominal process parameters, $K = 1$, $D = 1$, $T = 1$.

The PI controller is given and parameterized as follows:
\begin{equation*}
C_\mathrm{PI}(s) = K_\mathrm{P} + \D\frac{K_\mathrm{I}}{s}
\quad \text{with} \quad
K_\mathrm{I} = \frac{2.55}{K \cdot T_\mathrm{settle}} = 0.51
\quad \text{and} \quad
K_\mathrm{P} = K_\mathrm{I} \cdot 1.5 \cdot D \cdot T = 0.765
\end{equation*}

\begin{figure}[t!]
 \centering%
 \subfloat[]{%
 \includegraphics[width=0.32\linewidth]{adrc2_pt2_K_y.pdf}%
 }
 \subfloat[]{%
 \includegraphics[width=0.32\linewidth]{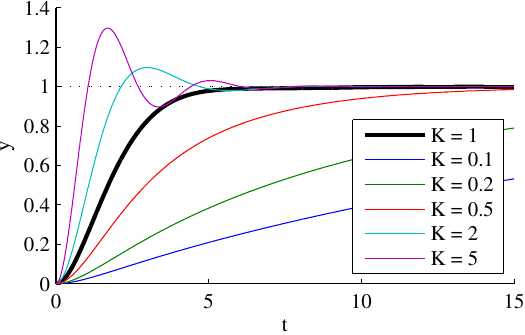}%
 }
 \subfloat[]{%
 \includegraphics[width=0.32\linewidth]{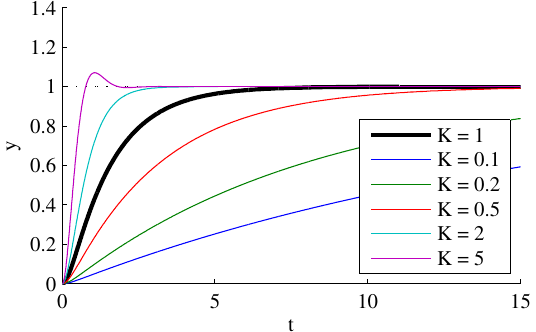}%
 }
 \\
 \subfloat[]{%
 \includegraphics[width=0.32\linewidth]{adrc2_pt2_D_y.pdf}%
 }
 \subfloat[]{%
 \includegraphics[width=0.32\linewidth]{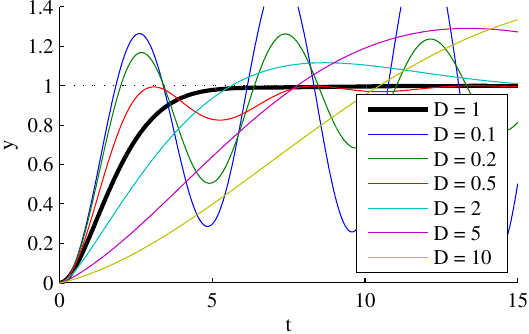}%
 }
 \subfloat[]{%
 \includegraphics[width=0.32\linewidth]{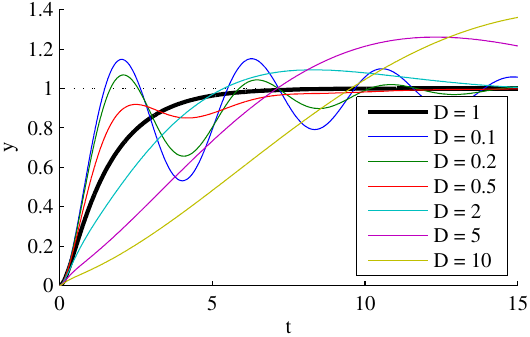}%
 }
 \\
 \subfloat[]{%
 \includegraphics[width=0.32\linewidth]{adrc2_pt2_T_y.pdf}%
 }
 \subfloat[]{%
 \includegraphics[width=0.32\linewidth]{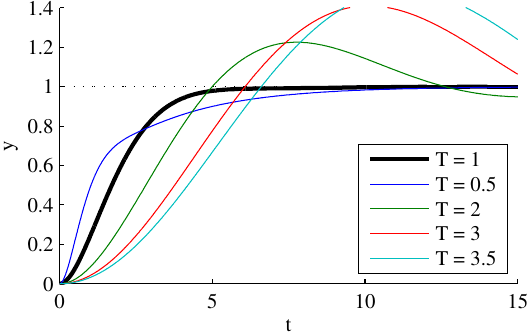}%
 }
 \subfloat[]{%
 \includegraphics[width=0.32\linewidth]{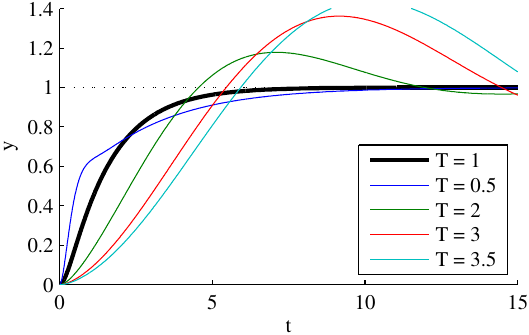}%
 }
 \caption{Experiment~\ref{sec:exp_adrc2_pid}: comparison of second-order ADRC, PI and PID controllers faced with varying process parameters, $K$, $D$ and $T$. Nominal process and ADRC parameters are as throughout \refSec{sec:exp_adrc2}. For each combination, the closed loop step response is shown. (\textbf{a}) Variation of $K$, ADRC; (\textbf{b}) Variation of $K$, PI; (\textbf{c}) Variation of $K$, PID; (\textbf{d}) Variation of $D$, ADRC; (\textbf{e}) Variation of $D$, PI; (\textbf{f}) Variation of $D$, PID; (\textbf{g}) Variation of $T$, ADRC; (\textbf{h}) Variation of $T$, PI; (\textbf{i}) Variation of $T$, PID.}
 \label{fig:adrc2_pt2_pid}
\end{figure}

For the PID controller, we will employ a $\mathrm{PIDT_1}$-type controller in two-pole-two-zero form, designed such that the zeros of the controller cancel the process poles. The controller gain is chosen to match the closed loop dynamics to the PI and ADRC case:
\begin{equation*}
C_\mathrm{PID}(s) = K_\mathrm{I} \cdot \D\frac{ (1 + T_\mathrm{Z1} \cdot s) (1 + T_\mathrm{Z2} \cdot s)}{s \cdot (1 + T_1 \cdot s)}
\quad \text{with} \quad
K_\mathrm{I} = \frac{3}{K \cdot T_\mathrm{settle}} = 0.6
,\quad
T_\mathrm{Z1/2} = T = 1
\quad \text{and} \quad
T_1 = 0.2
\end{equation*}

The simulation results are compiled in Figure \ref{fig:adrc2_pt2_pid}, where each column represents one of the three controller types and each row, a different process parameter being varied. To summarize, it has to be stated that in each case, the ADRC approach surpasses the results of PI and PID control by a large margin with respect to sensitivity towards parameter variations, with a slight advantage of PID compared to PI~control.


\begin{figure}[t]
 \centering%
 \subfloat[]{%
 \includegraphics[width=0.48\linewidth]{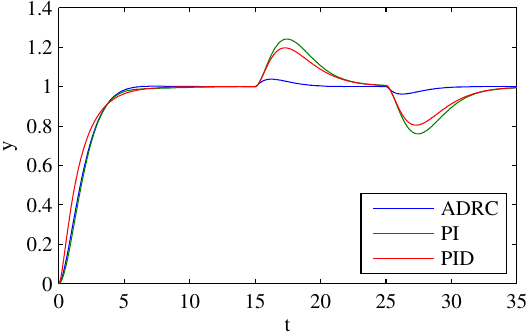}%
 }
 \subfloat[]{%
 \includegraphics[width=0.48\linewidth]{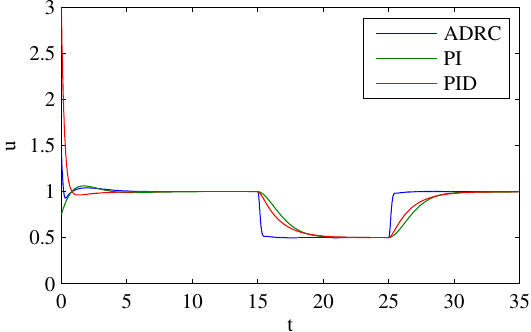}%
 }
 \caption{Experiment~\ref{sec:exp_adrc2_disturbance}: comparison of disturbance rejection behavior (second-order ADRC \textit{vs}.\ PI \textit{vs}.\ PID controller). Process and ADRC parameters are as throughout \refSec{sec:exp_adrc2}. Input disturbance, $d = 0.5$, effective from $t = 15$ until $t = 25$. (\textbf{a}) Step response and reaction on disturbance; (\textbf{b}) Controller output $u$ for (\textbf{a}).}
 \label{fig:adrc2_pt2_z}
\end{figure}

\subsubsection{Disturbance Rejection of ADRC, PI and PID}
\label{sec:exp_adrc2_disturbance}

As our final experiment for the second-order case, we will evaluate the disturbance rejection capabilities of ADRC, PI and PID control using the controller settings from the previous section. All of them are tuned for the same closed loop response.

For each control loop, an input disturbance, $d = 0.5$, is applied during a ten-second interval after reaching steady state; compare Figure \ref{fig:adrc2_pt2_z}. One can easily recognize that ADRC compensates the disturbance much faster, such that its effect on the control variable remains very low. While both PI and PID controllers could also be tuned more aggressively for a similar disturbance rejection behavior, one would have to employ additional measures, such as setpoint filters, to retain a non-oscillating reference tracking behavior.


\section{Discrete Time ADRC}
\label{sec:adrc_discrete}

Practical implementations of a controller with a state observer, such as the ADRC approach, will most likely be done in discrete time form, e.g.\ employing a microcontroller. Since the actual control law for linear ADRC is based on proportional state feedback, a discrete time version can already be obtained by only discretizing the extended state observer, which will be done in Section \ref{sec:adrc_discrete_eso}. The quasi-continuous approach will be valid only for sufficiently fast sampling, of course. Otherwise, the state feedback should be designed explicitly for a discretized process model incorporating sampling delays.

In Section \ref{sec:adrc_discrete_sim}, simulative experiments will be carried out in order to visually assess the influence of the discretisation process and measurement noise on the control loop performance.


\subsection{Discretisation of the State Observer}
\label{sec:adrc_discrete_eso}

For a system without a feed-through term, the standard approach for a discrete-time observer is:
\begin{equation}
\label{eqn:prediction_observer}
\Vector{\hat{x}}(k+1) = \Matrix{A}_\mathrm{d} \cdot \Vector{\hat{x}}(k) + \Matrix{B}_\mathrm{d} \cdot \Vector{u}(k) + \Matrix{L}_\mathrm{p} \cdot \left( y(k) - \Matrix{C}_\mathrm{d} \cdot \Vector{\hat{x}}(k) \right)
\end{equation}

If we look at the equation for the estimation error of Equation \refEq{eqn:prediction_observer}, we can see that---just as in the continuous case---the dynamics of the error decay are determined by the matrix $\left( \Matrix{A}_\mathrm{d} - \Matrix{L}_\mathrm{p} \cdot \Matrix{C}_\mathrm{d} \right)$:
\begin{equation*}
\Vector{e}(k+1)
= \Vector{x}(k+1) - \Vector{\hat{x}}(k+1)
= \left( \Matrix{A}_\mathrm{d} - \Matrix{L}_\mathrm{p} \cdot \Matrix{C}_\mathrm{d} \right) \cdot \left( \Vector{x}(k) - \Vector{\hat{x}}(k) \right)
\end{equation*}

Since the observer gains in $\Matrix{L}_\mathrm{p}$ influence the pole placement for the matrix $\left( \Matrix{A}_\mathrm{d} - \Matrix{L}_\mathrm{p} \cdot \Matrix{C}_\mathrm{d} \right)$, they can be chosen such that the estimation settles within a desired time.

Here, $\Matrix{A}_\mathrm{d}$, $\Matrix{B}_\mathrm{d}$ and $\Matrix{C}_\mathrm{d}$ refer to time-discrete versions of the respective matrices in the state space process models Equations \refEq{eqn:pt1_with_disturbance_ss} and \refEq{eqn:pt2_with_disturbance_ss} obtained by a discretisation method. Since there is no matrix $\Matrix{D}_\mathrm{d}$ in the observer equations, the discretisation of the model must deliver $\Matrix{D}_\mathrm{d} = 0$, for example, via zero order hold (ZOH) sampling. This observer approach is also being called ``prediction observer'', since the current measurement, $y(k)$, will be used as a correction of the estimate only in the subsequent time step, $\Vector{\hat{x}}(k+1)$.

In order to reduce unnecessary time delays (which may even destabilize the control loop), it is advisable \cite{Miklosovic:2006} to employ a different observer approach called ``filtered observer'' or ``current observer''~\cite{Franklin:1997}. The underlying idea is---similar to \Person{Kalman} filtering---to split the update equation in two steps, namely a prediction step to predict $\Vector{\hat{x}}(k|k-1)$ based on measurements of the previous time step, $k-1$, and a correction step to obtain the final estimate, $\Vector{\hat{x}}(k|k)$, incorporating the most recent measurement, $y(k)$:
\begin{equation*}
\Vector{\hat{x}}(k|k-1)
= \Matrix{A}_\mathrm{d} \cdot \Vector{\hat{x}}(k-1|k-1) + \Matrix{B}_\mathrm{d} \cdot \Vector{u}(k-1)
\quad \text{(prediction)}
\end{equation*}
\begin{equation*}
\Vector{\hat{x}}(k|k)
= \Vector{\hat{x}}(k|k-1) + \Matrix{L}_\mathrm{c} \cdot \left( y(k) - \Matrix{C}_\mathrm{d} \cdot \Vector{\hat{x}}(k|k-1) \right)
\quad \text{(correction)}
\end{equation*}

If we put the prediction into the correction equation and abbreviate $\Vector{\hat{x}}(k|k) = \Vector{\hat{x}}(k)$, we obtain one update equation for the observer:
\begin{equation}
\label{eqn:current_observer}
\Vector{\hat{x}}(k)
= \left( \Matrix{A}_\mathrm{d} - \Matrix{L}_\mathrm{c} \cdot \Matrix{C}_\mathrm{d} \cdot \Matrix{A}_\mathrm{d} \right) \cdot \Vector{\hat{x}}(k-1) + \left( \Matrix{B}_\mathrm{d} - \Matrix{L}_\mathrm{c} \cdot \Matrix{C}_\mathrm{d} \cdot \Matrix{B}_\mathrm{d} \right) \cdot u(k-1) + \Matrix{L}_\mathrm{c} \cdot y(k)
\end{equation}

From the estimation error, we can see that, in contrast to the prediction observer, the error dynamics are determined by the matrix $\left( \Matrix{A}_\mathrm{d} - \Matrix{L}_\mathrm{c} \cdot \Matrix{C}_\mathrm{d} \cdot \Matrix{A}_\mathrm{d} \right)$:
\begin{equation*}
\Vector{e}(k+1)
= \Vector{x}(k+1) - \Vector{\hat{x}}(k+1)
= \left( \Matrix{A}_\mathrm{d} - \Matrix{L}_\mathrm{c} \cdot \Matrix{C}_\mathrm{d} \cdot \Matrix{A}_\mathrm{d} \right) \cdot \left( \Vector{x}(k) - \Vector{\hat{x}}(k) \right)
\end{equation*}

When computing the observer gains in $\Matrix{L}_\mathrm{c}$, this has to be taken into account, \textit{i.e}., one must choose $\Matrix{L}_\mathrm{c}$, such that the eigenvalues of $\left( \Matrix{A}_\mathrm{d} - \Matrix{L}_\mathrm{c} \cdot \Matrix{C}_\mathrm{d} \cdot \Matrix{A}_\mathrm{d} \right)$ match the desired observer pole locations.

The discrete time versions of the matrices $\Matrix{A}$, $\Matrix{B}$ and $\Matrix{C}$ from the the state space process models, which are necessary for the observer equations, can be obtained by ZOH discretisation \cite{Franklin:1997}:
\begin{equation}
\label{eqn:zoh_transform}
\Matrix{A}_\mathrm{d}
= \Matrix{I} + \sum_{i = 1}^\infty \D\frac{\Matrix{A}^i \cdot T^i_\mathrm{sample}}{i!},\quad
\Matrix{B}_\mathrm{d}
= \left( \sum_{i = 1}^\infty \D\frac{\Matrix{A}^{i-1} \cdot T^i_\mathrm{sample}}{i!} \right) \cdot \Matrix{B},\quad
\Matrix{C}_\mathrm{d} = \Matrix{C},\quad
\Matrix{D}_\mathrm{d} = \Matrix{D}
\end{equation}

For the first-order process in Equation \refEq{eqn:pt1_with_disturbance_ss}, $\Matrix{A}$ and $\Matrix{B}$ are being discretized as follows, while
$\Matrix{C}_\mathrm{d} = \Matrix{C} = \begin{pmatrix}
 1 & 0
\end{pmatrix}$
and
$\Matrix{D}_\mathrm{d} = \Matrix{D} = 0$ remain unchanged:
\begin{equation*}
\Matrix{A} =
\begin{pmatrix}
0 & 1 \\
0 & 0
\end{pmatrix}
,\quad
\Matrix{B} =
\begin{pmatrix}
b_0 \\ 0
\end{pmatrix}
\quad\text{gives}\quad
\Matrix{A}_\mathrm{d} =
\begin{pmatrix}
1 & T_\mathrm{sample} \\
0 & 1
\end{pmatrix}
,\quad
\Matrix{B}_\mathrm{d} =
\begin{pmatrix}
b_0 \cdot T_\mathrm{sample} \\ 0
\end{pmatrix}
\end{equation*}

This can be computed via Equation \refEq{eqn:zoh_transform} easily, since $\Matrix{A}^i = \Matrix{0}$ for $i \ge 2$. Following the same procedure for the second-order case in Equation \refEq{eqn:pt2_with_disturbance_ss}, $\Matrix{C}_\mathrm{d} = \Matrix{C} = \begin{pmatrix}
 1 & 0 & 0
\end{pmatrix}$
and
$\Matrix{D}_\mathrm{d} = \Matrix{D} = 0$
remain unchanged, and one obtains for $\Matrix{A}_\mathrm{d}$ and $\Matrix{B}_\mathrm{d}$ (since $\Matrix{A}^i = \Matrix{0}$ for $i \ge 3$):
\begin{equation*}
\Matrix{A} =
\begin{pmatrix}
0 & 1 & 0 \\
0 & 0 & 1 \\
0 & 0 & 0
\end{pmatrix}
,\quad
\Matrix{B} =
\begin{pmatrix}
0 \\ b_0 \\ 0
\end{pmatrix}
\quad\text{gives}\quad
\Matrix{A}_\mathrm{d} =
\begin{pmatrix}
1 & T_\mathrm{sample} & T^2_\mathrm{sample}/2 \\
0 & 1 & T_\mathrm{sample} \\
0 & 0 & 1
\end{pmatrix}
,\quad
\Matrix{B}_\mathrm{d} =
\begin{pmatrix}
b_0 \cdot T^2_\mathrm{sample}/2 \\ b_0 \cdot T_\mathrm{sample} \\ 0
\end{pmatrix}
\end{equation*}

Now, one can compute the observer gain, $\Matrix{L}_\mathrm{c} = \begin{pmatrix} l_1 & l_2 \end{pmatrix}^\mathrm{T}$ (first-order case) or $\Matrix{L}_\mathrm{c} = \begin{pmatrix} l_1 & l_2 & l_3 \end{pmatrix}^\mathrm{T}$ (for the second-order ADRC) to obtain the desired observer dynamics. The desired pole locations can, in a first step, be formulated in the $s$-plane, as in Sections \ref{sec:adrc1} and \ref{sec:adrc2}, and then be mapped to the $z$-plane: $z^\mathrm{ESO} = \mathrm{e}^{s^\mathrm{ESO} \cdot T_\mathrm{sample}}$. The computation of $\Matrix{L}_\mathrm{c}$ may either be done numerically or analytically. To demonstrate this, we will derive the equations for the first-order case with one common pole location, $z^\mathrm{ESO}$, for both poles:
\begin{align*}
\det \left( z \Matrix{I} - \left( \Matrix{A}_\mathrm{d} - \Matrix{L}_\mathrm{c} \cdot \Matrix{C}_\mathrm{d} \cdot \Matrix{A}_\mathrm{d} \right) \right)
& \stackrel{!}{=} \left( z - z^\mathrm{ESO} \right)^2
\\
\det \begin{pmatrix}
z + l_1 - 1 & l_1 \cdot T_\mathrm{sample} - T_\mathrm{sample} \\
l_2 & z + l_2 \cdot T_\mathrm{sample} - 1
\end{pmatrix}
& \stackrel{!}{=} z^2 - 2 \cdot z^\mathrm{ESO} \cdot z + \left( z^\mathrm{ESO} \right)^2
\\
z^2 + \left( l_1 + l_2 \cdot T_\mathrm{sample} -2 \right) \cdot z + \left( 1 - l_1 \right)
& \stackrel{!}{=} z^2 - 2 \cdot z^\mathrm{ESO} \cdot z + \left( z^\mathrm{ESO} \right)^2
\end{align*}

By comparing the coefficients, one obtains for $l_1$ and $l_2$ (solutions for higher-order observers can be found in \cite{Miklosovic:2006}):%
\endnote[\textcolor{black!50}{\textdagger}]{%
    In this updated version of the preprint, equation (\ref{eqn:dt_observer_gains}) and its derivation have been corrected. The correct result for the observer gain $l_2$ is
    $l_2 = \frac{1}{T_\mathrm{sample}} \cdot \left( 1 - z^\mathrm{ESO} \right)^2$
    \quad
    instead of the incorrect equation
    \quad
    $l_2 = \frac{1}{T_\mathrm{sample}} \cdot \left( 1 - \left( z^\mathrm{ESO} \right)^2 \right)$
    \quad
    previously reported in the original publication, which is available at
    \href{http://dx.doi.org/10.3390/electronics2030246}{doi:10.3390/electronics2030246}.
    The author would like to thank Rafa\l{} Mado\'{n}ski for reporting this issue.
}
\begin{equation}
l_1 = 1 - \left( z^\mathrm{ESO} \right)^2,\quad
l_2 = \frac{1}{T_\mathrm{sample}} \cdot \left( 1 - z^\mathrm{ESO} \right)^2
\label{eqn:dt_observer_gains}
\end{equation}

As in the continuous time case in Section \ref{sec:adrc}, the implementation and design of a discrete time linear ADRC can be summarized in four steps, where the first-order and second-order case are distinguished by (a) and (b):
\begin{enumerate}
\item
\emph{Modeling:}

\begin{enumerate}
\item
For a process with dominating first-order behavior, $P(s) = \D\frac{K}{Ts + 1}$, one should provide an estimate $b_0 \approx \frac{K}{T}$.

\item
For a second-order process, $P(s) = \D\frac{K}{T^2 s^2 + 2DTs + 1}$, an approximate value $b_0 \approx \frac{K}{T^2}$ is~sufficient.
\end{enumerate}

\item
\emph{Control structure:}

Implement a discrete-time observer with two state variables, $\Vector{\hat{x}}(k) = \begin{pmatrix} \hat{x}_1(k) & \hat{x}_2(k) \end{pmatrix}^\mathrm{T}$ (for the first-order case) or three state variables $\Vector{\hat{x}}(k) = \begin{pmatrix} \hat{x}_1(k) & \hat{x}_2(k) & \hat{x}_3(k) \end{pmatrix}^\mathrm{T}$ (for the second-order case), as follows:

$\Vector{\hat{x}}(k)
= \left( \Matrix{A}_\mathrm{d} - \Matrix{L}_\mathrm{c} \cdot \Matrix{C}_\mathrm{d} \cdot \Matrix{A}_\mathrm{d} \right) \cdot \Vector{\hat{x}}(k-1) + \left( \Matrix{B}_\mathrm{d} - \Matrix{L}_\mathrm{c} \cdot \Matrix{C}_\mathrm{d} \cdot \Matrix{B}_\mathrm{d} \right) \cdot u(k-1) + \Matrix{L}_\mathrm{c} \cdot y(k)
$
\smallskip

Using the estimated state variables, implement a proportional state feedback controller:
\begin{enumerate}
\item
$u(k)
= \D\frac{K_\mathrm{P} \cdot \left( r(k) - \hat{y}(k) \right) - \hat{f}(k)}{b_0}
= \frac{K_\mathrm{P} \cdot \left( r(k) - \hat{x}_1(k) \right) - \hat{x}_2(k)}{b_0}
$

\item
$
u(k) = \D\frac{\left( K_\mathrm{P} \cdot \left( r(k) - \hat{y}(k) \right) - K_\mathrm{D} \cdot \dot{\hat{y}}(k) \right) - \hat{f}(k)}{b_0}
= \D\frac{\left( K_\mathrm{P} \cdot \left( r(k) - \hat{x}_1(k) \right) - K_\mathrm{D} \cdot \hat{x}_2(k) \right) - \hat{x}_3(k)}{b_0}
$
\end{enumerate}

\item
\emph{Closed loop dynamics:}

Choose the controller parameters, e.g.\ via pole placement according to a desired settling time:
\begin{enumerate}
\item
$K_\mathrm{P} = -s^\mathrm{CL}
\quad \text{with} \quad
s^\mathrm{CL} \approx -\D\frac{4}{T_\mathrm{settle}}$

\item
$
K_\mathrm{P} = \left(s^\mathrm{CL}\right)^2
,\quad
K_\mathrm{D} = -2 \cdot s^\mathrm{CL}
\quad \text{with} \quad
s^\mathrm{CL} \approx -\D\frac{6}{T_\mathrm{settle}}
$

\end{enumerate}

\item
\emph{Observer dynamics:}

Place the desired observer poles far enough left of the closed loop poles in the $s$-plane, e.g.\ by a common location, $s^\mathrm{ESO} = k_\mathrm{ESO} \cdot s^\mathrm{CL} \approx (3 \ldots 10) \cdot s^\mathrm{CL}$. Map to the $z$-plane via $z^\mathrm{ESO} = \mathrm{e}^{s^\mathrm{ESO} \cdot T_\mathrm{sample}}$ and compute the necessary observer gains:
\begin{enumerate}
\item
$l_1 = 1 - \left(z^\mathrm{ESO}\right)^2$,\quad
$l_2 = \frac{1}{T_\mathrm{sample}} \cdot \left(1 - z^\mathrm{ESO}\right)^2$

\item
$l_1 = 1 - \left(z^\mathrm{ESO}\right)^3$,\quad
$l_2 = \frac{3}{2 \cdot T_\mathrm{sample}} \cdot \left(1 - z^\mathrm{ESO}\right)^2 \cdot \left(1 + z^\mathrm{ESO}\right)$,\quad
$l_3 = \frac{1}{T^2_\mathrm{sample}} \cdot \left(1 - z^\mathrm{ESO}\right)^3$

\end{enumerate}

\end{enumerate}


\subsection{Simulative Experiments}
\label{sec:adrc_discrete_sim}

As in the continuous-time experiments in Section \ref{sec:exp_adrc1}, a first-order process will be examined in this~section:
\begin{equation*}
P(s) = \D\frac{y(s)}{u(s)} = \D\frac{K}{Ts + 1}
\quad \text{with} \quad
K = 1 \quad \text{and} \quad T = 1
\end{equation*}

Unless otherwise noted in individual experiments, the ADRC will be designed assuming perfect knowledge, $b_0 = K/T = 1$, with $k_\mathrm{ESO} = 5$, and a desired 2\% settling time, $T_\mathrm{settle} = 1$. The discretisation will be based on a sampling time $T_\mathrm{sample} = 0.01$. Gaussian measurement noise will be added with a variance $\sigma^2_\mathrm{noise} = 0.0001$.


\subsubsection{Effect of Sample Time}
\label{sec:exp_adrc1d_Tsample}

When choosing the sampling time for a discrete-time implementation of a controller with an observer, one has to consider not only the process dynamics, but also the dynamics of the observer. If, on the other hand, restrictions to sample times are present, the dynamics of the observer will be limited. In the case of ADRC, this would mean that the desired behavior of the process may not be achieved under all~circumstances.

\begin{figure}[t!]
 \centering%
 \subfloat[]{%
 \includegraphics[width=0.48\linewidth]{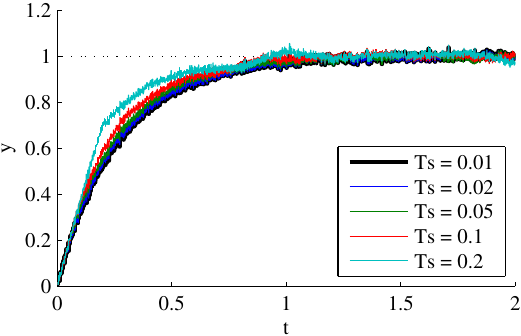}%
 }
 \subfloat[]{%
 \includegraphics[width=0.48\linewidth]{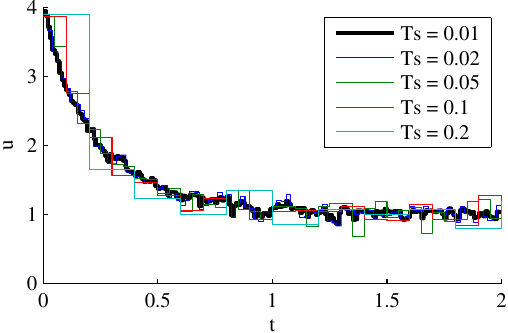}%
 }
 \caption{Experiment~\ref{sec:exp_adrc1d_Tsample}: influence of different sample time settings on discrete-time ADRC. Measurement noise variance: $\sigma^2_\mathrm{noise} = 0.0001$. Observer pole placement: $k_\mathrm{ESO} = 5$. (\textbf{a}) Variation of $T_\mathrm{sample}$, closed loop step response; (\textbf{b}) Controller output $u$ for (\textbf{a}).}
 \label{fig:adrc1d_Tsample}
\end{figure}

In Figure \ref{fig:adrc1d_Tsample}, simulations were performed using sample times for ADRC ranging from $T_\mathrm{sample} = 0.01$ to $T_\mathrm{sample} = 0.20$. As expected, oscillations in the controller output are increasingly visible for large sample times. Consequently, the closed loop dynamics differ from the desired first-order behavior as the sampling interval becomes too coarse.


\begin{figure}[t!]
 \centering%
 \subfloat[]{%
 \includegraphics[width=0.48\linewidth]{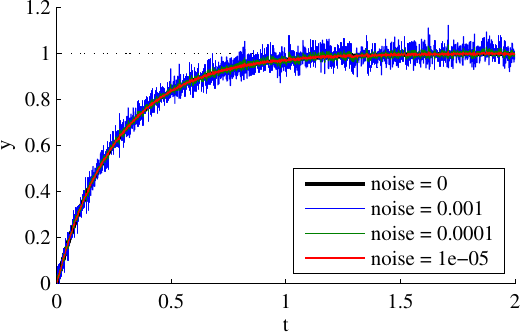}%
 }
 \subfloat[]{%
 \includegraphics[width=0.48\linewidth]{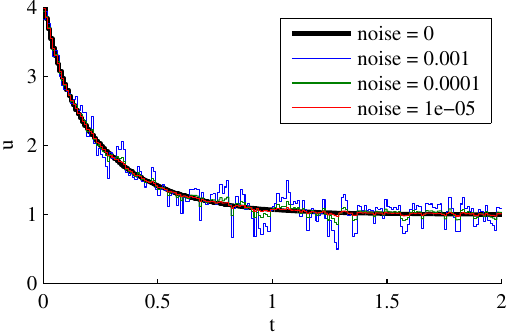}%
 }
 \caption{Experiment~\ref{sec:exp_adrc1d_noise}: influence of different levels of measurement noise on discrete-time ADRC. Sample time: $T_\mathrm{sample} = 0.01$. Observer pole placement: $k_\mathrm{ESO} = 5$. (\textbf{a}) Variation of $\sigma^2_\mathrm{noise}$, closed loop step response; (\textbf{b}) Controller output $u$ for (\textbf{a}).}
 \label{fig:adrc1d_noise}
\end{figure}

\subsubsection{Effect of Measurement Noise}
\label{sec:exp_adrc1d_noise}

While state feedback controllers based on observers rather than direct measurements are less susceptible to measurement noise, oscillations in the controller output will still occur with higher noise levels. In the simulations presented in Figure \ref{fig:adrc1d_noise}, normally distributed noise with increasing variance ranging from $\sigma^2_\mathrm{noise} = 0$ to $\sigma^2_\mathrm{noise} = 0.001$ was added to the process output. The results correspond to these expectations. The effect of measurement noise on oscillations in the controller output, however, may be mitigated by designing an observer with slower dynamics, which will be demonstrated in the following experiment.


\begin{figure}[t]
 \centering%
 \subfloat[]{%
 \includegraphics[width=0.48\linewidth]{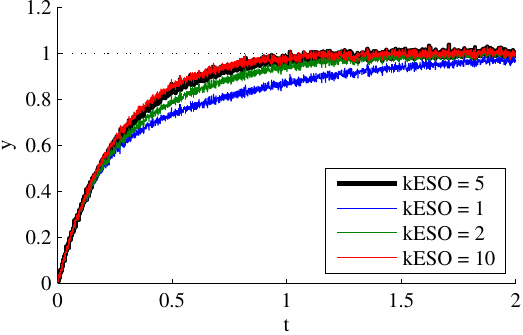}%
 }
 \subfloat[]{%
 \includegraphics[width=0.48\linewidth]{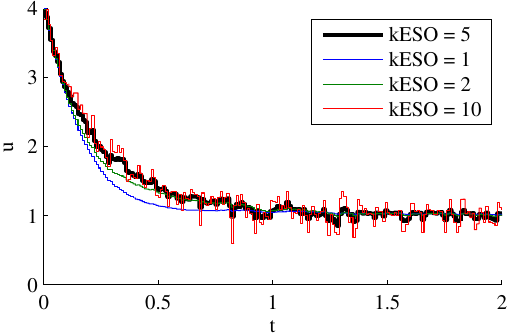}%
 }
 \caption{Experiment~\ref{sec:exp_adrc1d_kESO}: influence of observer pole placement via $k_\mathrm{ESO}$ on discrete-time ADRC. Sample time: $T_\mathrm{sample} = 0.01$. Measurement noise variance: $\sigma^2_\mathrm{noise} = 0.0001$. (\textbf{a}) Variation of $k_\mathrm{ESO}$, closed loop step response; (\textbf{b}) Controller output $u$ for (\textbf{a}).}
 \label{fig:adrc1d_kESO}
\end{figure}

\subsubsection{Effect of Observer Pole Locations}
\label{sec:exp_adrc1d_kESO}

In the simulations with varying observer dynamics presented in Figure \ref{fig:adrc1d_kESO}, the typical trade-off between fast setpoint tracking and noise rejection becomes visible. If the observer poles are placed far enough left of the closed loop poles in the $s$-plane via $k_\mathrm{ESO}$, the closed loop dynamics can adopt the desired first-order behavior, but the control variable becomes more sensitive to measurement noise and increasingly exhibits unwanted oscillations. On the other hand, placing observer poles near the closed loop poles provides good measures to suppress the effect of measurement noise on the control action, but the closed loop dynamics will suffer, especially under process parameter changes, \textit{cf}.\ Sections \ref{sec:exp_adrc1_eso} and \ref{sec:exp_adrc2_eso}.


\section{Optimized Discrete-Time Implementation}

In a practical discrete-time implementation with tight timing constraints, the lag between sensor input and controller output should be as small as possible, which means that one should try to reduce the computational effort for a controller. Due to the observer-based approach, ADRC has a bigger computational footprint than a classical PID-type controller. In this section, we will try to reduce the number of computations, on the one hand, and present an implementation focused on low input-output delay, on the other hand.

\subsection{State Variable Transformation}

We start with the discrete-time observer from Equation (\ref{eqn:current_observer}) in the following abbreviated form:
\begin{equation*}
\Vector{\hat{x}}(k)
= \Matrix{A}_\mathrm{ESO} \cdot \Vector{\hat{x}}(k-1) + \Matrix{B}_\mathrm{ESO} \cdot u(k-1) + \Matrix{L}_\mathrm{ESO} \cdot y(k)
\end{equation*}
and with a control law for the first-order or second-order case:
\begin{equation*}
u(k)
= \D\frac{K_\mathrm{P} \cdot \left( r(k) - \hat{x}_1(k) \right) - \hat{x}_2(k)}{b_0}
\quad\text{or}\quad
u(k)
= \D\frac{\left( K_\mathrm{P} \cdot \left( r(k) - \hat{x}_1(k) \right) - K_\mathrm{D} \cdot \hat{x}_2(k) \right) - \hat{x}_3(k)}{b_0}
\end{equation*}

One can simplify the controller structure by scaling the outputs of the observer, such that the multiplications by $b_0$, $K_\mathrm{P}$ and $K_\mathrm{D}$ (the latter only for the second-order case) can be omitted. The desired scaling of the new state variables, $\widetilde{x}_i$, is achieved via:
\begin{itemize}
\item[(a)]
$\D\tilde{x}_1 = \frac{K_\mathrm{P}}{b_0} \cdot \hat{x}_1$, \quad
$\D\tilde{x}_2 = \frac{1}{b_0} \cdot \hat{x}_2$
\quad (for the first-order case)
\medskip

\item[(b)]
$\D\tilde{x}_1 = \frac{K_\mathrm{P}}{b_0} \cdot \hat{x}_1$, \quad
$\D\tilde{x}_2 = \frac{K_\mathrm{D}}{b_0} \cdot \hat{x}_2$, \quad
$\D\tilde{x}_3 = \frac{1}{b_0} \cdot \hat{x}_3$
\quad (for the second-order case)
\end{itemize}

\begin{figure}
 \centering%
 \subfloat[]{%
 \centering\includegraphics{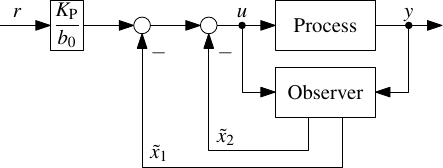}%
 }
 \quad
 \subfloat[]{%
 \centering\includegraphics{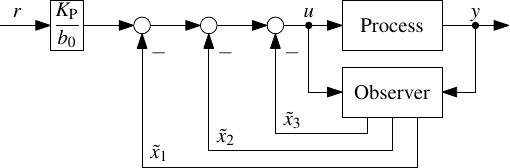}%
 }
 \caption{Control loop with modified ADRC structure and transformed state variables, $\tilde{x}$. (\textbf{a}) Modified first-order structure; (\textbf{b}) Modified second-order structure.}
 \label{fig:adrc_structure_modified}
\end{figure}

The ADRC structure can then be modified as given in Figure \ref{fig:adrc_structure_modified}. Generally speaking, we perform a coordinate transformation from the old estimated variables $\Vector{\hat{x}}$ to $\Vector{\tilde{x}}$ by means of a transformation matrix:
\begin{equation}
\Matrix{T}^{-1}
=
\D\frac{1}{b_0} \cdot
\begin{pmatrix}
K_\mathrm{P} & 0 \\
0 & 1 \\
\end{pmatrix}
\quad \text{(1}^\mathrm{st} \text{ order ADRC)}
\quad \text{or} \quad
\Matrix{T}^{-1}
=
\D\frac{1}{b_0} \cdot
\begin{pmatrix}
K_\mathrm{P} & 0 & 0 \\
0 & K_\mathrm{D} & 0 \\
0 & 0 & 1 \\
\end{pmatrix}
\quad \text{(2}^\mathrm{nd} \text{ order)}
\end{equation}

Using this matrix $\Matrix{T}$, the state space equations for the extended state observer must be transformed in order to obtain an ESO working with our new state variables, $\Vector{\tilde{x}} = \Matrix{T}^{-1} \cdot \Vector{\hat{x}}$:
\begin{equation}
\Matrix{\widetilde{A}}_\mathrm{ESO} = \Matrix{T}^{-1} \cdot \Matrix{A}_\mathrm{ESO} \cdot \Matrix{T}, \quad
\Matrix{\widetilde{B}}_\mathrm{ESO} = \Matrix{T}^{-1} \cdot \Matrix{B}_\mathrm{ESO}, \quad
\Matrix{\widetilde{L}}_\mathrm{ESO} = \Matrix{T}^{-1} \cdot \Matrix{L}_\mathrm{ESO}
\end{equation}

In this manner, multiplications of the state variables by $b_0$, $K_\mathrm{P}$ and $K_\mathrm{D}$ can be avoided. Provided the reference, $r(k)$, does not change at each point in time, one can precompute $\frac{K_\mathrm{P}}{b_0} \cdot r$, as well. The control laws are being simplified to the following form:
\begin{equation*}
u(k)
= \D\frac{K_\mathrm{P}}{b_0} \cdot r(k) - \hat{x}_1(k) - \hat{x}_2(k)
\quad\text{or}\quad
u(k)
= \D\frac{K_\mathrm{P}}{b_0} \cdot r(k) - \tilde{x}_1(k) - \tilde{x}_2(k) - \tilde{x}_3(k)
\end{equation*}


\subsection{Minimizing Latency by Precomputation}

In an application with fixed sampling frequency, the controller performance can be improved by minimizing the latency between acquisition of input signals and output of the updated controller action within a sampling period. This input-output lag is not necessarily dependent on the overall number of computations of a control law, but rather on the computations necessary to deliver the controller output, with the computations left being performed after that during the remaining time of the sampling period.

We will optimize the discrete-time implementation of ADRC in this regard and, subsequently, derive further equations in detail for the second-order case. Following the ideas of the previous section, the estimated state variables must be updated according to Equation (\ref{eqn:latency_observer}) and used to compute the controller output in Equation (\ref{eqn:latency_controller}), at each point in time $k$:
\begin{equation}
\label{eqn:latency_observer}
\Vector{\tilde{x}}(k)
= \Matrix{\widetilde{A}}_\mathrm{ESO} \cdot \Vector{\tilde{x}}(k-1) + \Matrix{\widetilde{B}_}\mathrm{ESO} \cdot u(k-1) + \Matrix{\widetilde{L}}_\mathrm{ESO} \cdot y(k)
\end{equation}
\begin{equation*}
\text{with}\quad
\Vector{\tilde{x}} =
\begin{pmatrix}
\tilde{x}_1 \\ \tilde{x}_2 \\ \tilde{x}_3
\end{pmatrix}
,\quad
\Matrix{\widetilde{A}}_\mathrm{ESO}
=
\begin{pmatrix}
\widetilde{a}_{11} & \widetilde{a}_{12} & \widetilde{a}_{13} \\
\widetilde{a}_{21} & \widetilde{a}_{22} & \widetilde{a}_{33} \\
\widetilde{a}_{31} & \widetilde{a}_{32} & \widetilde{a}_{33} \\
\end{pmatrix}
,\quad
\Matrix{\widetilde{B}}_\mathrm{ESO}
=
\begin{pmatrix}
\widetilde{b}_1 \\ \widetilde{b}_2 \\ \widetilde{b}_3
\end{pmatrix}
,\quad
\Matrix{\widetilde{L}}_\mathrm{ESO}
=
\begin{pmatrix}
\tilde{l}_1 \\ \tilde{l}_2 \\ \tilde{l}_3
\end{pmatrix}
\end{equation*}

\begin{equation}
\label{eqn:latency_controller}
u(k)
= \D\frac{K_\mathrm{P}}{b_0} \cdot r(k) - \tilde{x}_1(k) - \tilde{x}_2(k) - \tilde{x}_3(k)
= \D\frac{K_\mathrm{P}}{b_0} \cdot r(k) -
\begin{pmatrix} 1 & 1 & 1 \end{pmatrix} \cdot \Vector{\tilde{x}}(k)
\end{equation}

When inserting Equation (\ref{eqn:latency_observer}) into Equation (\ref{eqn:latency_controller}), one can see that $u(k)$ depends both on variables from time $k$, $y(k)$ and $r(k)$, and from time $k-1$, $u(k-1)$ and $\Vector{\tilde{x}}(k-1)$. An idea for providing $u(k)$ with lower latency would therefore be to precompute all terms stemming from time $k-1$ already at $k-1$ and only update this precomputed value, $u(k|k-1)$, at time $k$ to obtain and output $u(k) = u(k|k)$, as soon as the necessary measurements become available. Further optimization is possible if the reference value, $r$, does not change rapidly, remains constant or is known in advance. We will denote this by $r(k+1|k)$, \textit{i.e}., $r(k+1)$ is already known at time $k$, for any of the reasons mentioned. Then, only the measured process output, $y(k)$, has to be included in the update step to obtain $u(k) = u(k|k)$:
\begin{equation}
\label{eqn:latency1}
u(k) = u(k|k-1) - \begin{pmatrix} 1 & 1 & 1 \end{pmatrix} \cdot \Matrix{\widetilde{L}}_\mathrm{ESO} \cdot y(k)
= u(k|k-1) - \left( \tilde{l}_1 + \tilde{l}_2 + \tilde{l}_3 \right) \cdot y(k)
\end{equation}

In Equation (\ref{eqn:latency1}), the sum $(\tilde{l}_1 + \tilde{l}_2 + \tilde{l}_3)$ can be precomputed as well, such that only one multiplication and one addition is necessary to calculate the controller output, $u(k)$. After that, the remaining time of the sampling period can be used to update the observer states via Equation (\ref{eqn:latency2}) and precompute the output via Equation (\ref{eqn:latency3}) for the next point in time, $k+1$:
\begin{equation}
\label{eqn:latency2}
\Vector{\tilde{x}}(k+1|k) = \Matrix{\widetilde{A}}_\mathrm{ESO} \cdot \underbrace{ \left( \Vector{\tilde{x}}(k|k-1) + \Matrix{\widetilde{L}}_\mathrm{ESO} \cdot y(k) \right) }_{ \Vector{\tilde{x}}(k) } + \Matrix{\widetilde{B}}_\mathrm{ESO} \cdot u(k) \\
\end{equation}
\begin{equation}
\label{eqn:latency3}
u(k+1|k) = \D\frac{K_\mathrm{P}}{b_0} \cdot r(k+1|k) - \begin{pmatrix} 1 & 1 & 1 \end{pmatrix} \cdot \Vector{\tilde{x}}(k+1|k)
\end{equation}

Note that the actual observer states, $\Vector{\tilde{x}}(k)$, are not explicitly present and updated in the equations anymore, since only the precomputed values, $\Vector{\tilde{x}}(k+1|k)$, are needed. To summarize, a latency-optimized discrete-time implementation has to perform the following four steps at each point in time $k$, with the new controller output, $u(k)$, being available already after the second~step:
\begin{enumerate}
\item
Acquire the current measurement of the process output, $y(k)$.

\item
Calculate and output $u(k)$ by Equation (\ref{eqn:latency1}).

\item
Update the internal observer states by Equation (\ref{eqn:latency2}).

\item
Precompute the controller output for $k+1$ using Equation (\ref{eqn:latency3}).

\end{enumerate}


\section{Conclusions}

By means of simulative experiments using generic first-order and second-order plants, it could be demonstrated that ADRC can be a powerful control tool. In this article, the linear case was examined. It can adapt even to heavily varying process parameters, and---in contrast to ``classical'' adaptive controllers---do so without having to maintain an explicit model of the process. Since only little knowledge about a process has to be provided along with the desired closed loop dynamics, the parameterization is easy for both continuous time and discrete time cases and, therefore, appealing to~practitioners.

For control problems with high dynamic requirements, an optimized formulation of the discrete time linear ADRC equations can be found, which enables the controller output to be computed with only one addition and one multiplication after the sensor input becomes available.

If ADRC can show its full potential in a specific application does, however, depend on the relation of process dynamics, observer dynamics, sampling time and measurement noise. In order to provide the adaptability, the observer has to be fast enough compared to the process and closed loop dynamics, on the one hand. On the other hand, the placement of the observer poles will be limited by sampling frequency and the increasing effects of noise on the control action. As long as a good compromise can be found in this regard, ADRC has to be considered as a strong and welcome alternative to solving practical control problems.


{\color{black!50}%
\printendnotes
}


\begin{thebibliography}{10}

\bibitem{Araki:2003}
Mituhiko Araki and Hidefumi Taguchi.
\newblock Two-degree-of-freedom {PID} controllers.
\newblock {\em International Journal of Control, Automation, and Systems},
  1(4):401--411, December 2003.

\bibitem{Canuto:2007}
Enrico Canuto.
\newblock {E}mbedded {M}odel {C}ontrol: {O}utline of the theory.
\newblock {\em ISA Transactions}, 46(3):363--377, 2007.
\newblock \href {http://dx.doi.org/10.1016/j.isatra.2007.01.006}
  {\path{doi:10.1016/j.isatra.2007.01.006}}.

\bibitem{Chen:2011}
Xing Chen, Donghai Li, Zhiqiang Gao, and Chuanfeng Wang.
\newblock Tuning method for second-order active disturbance rejection control.
\newblock In {\em Proceedings of the 30th Chinese Control Conference}, pages
  6322--6327, 2011.

\bibitem{Francis:1976}
Bruce~A. Francis and Walter~Murray Wonham.
\newblock The internal model principle of control theory.
\newblock {\em Automatica}, 12(5):457--465, 1976.
\newblock \href {http://dx.doi.org/10.1016/0005-1098(76)90006-6}
  {\path{doi:10.1016/0005-1098(76)90006-6}}.

\bibitem{Franklin:1997}
Gene~F. Franklin, Michael~L. Workman, and Dave Powell.
\newblock {\em Digital Control of Dynamic Systems}.
\newblock Addison-Wesley Longman Publishing, Boston, MA, USA, 3rd edition,
  1997.

\bibitem{Gao:2003}
Zhiqiang Gao.
\newblock Scaling and bandwidth-parameterization based controller tuning.
\newblock In {\em Proceedings of the 2003 American Control Conference}, pages
  4989--4996, 2003.
\newblock \href {http://dx.doi.org/10.1109/ACC.2003.1242516}
  {\path{doi:10.1109/ACC.2003.1242516}}.

\bibitem{Gao:2006}
Zhiqiang Gao.
\newblock Active disturbance rejection control: {A} paradigm shift in feedback
  control system design.
\newblock In {\em Proceedings of the 2006 American Control Conference}, pages
  2399--2405, 2006.
\newblock \href {http://dx.doi.org/10.1109/ACC.2006.1656579}
  {\path{doi:10.1109/ACC.2006.1656579}}.

\bibitem{Gao:2001}
Zhiqiang Gao, Yi~Huang, and Jingqing Han.
\newblock An alternative paradigm for control system design.
\newblock In {\em Proceedings of the 40th IEEE Conference on Decision and
  Control}, 2001.
\newblock \href {http://dx.doi.org/10.1109/CDC.2001.980926}
  {\path{doi:10.1109/CDC.2001.980926}}.

\bibitem{Han:2009}
Jingqing Han.
\newblock From {PID} to active disturbance rejection control.
\newblock {\em IEEE Transactions on Industrial Electronics}, 56(3):900--906,
  2009.
\newblock \href {http://dx.doi.org/10.1109/TIE.2008.2011621}
  {\path{doi:10.1109/TIE.2008.2011621}}.

\bibitem{Miklosovic:2006}
Robert Miklosovic, Aaron Radke, and Zhiqiang Gao.
\newblock Discrete implementation and generalization of the extended state
  observer.
\newblock In {\em Proceedings of the 2006 American Control Conference}, pages
  2209--2214, 2006.
\newblock \href {http://dx.doi.org/10.1109/ACC.2006.1656547}
  {\path{doi:10.1109/ACC.2006.1656547}}.

\bibitem{Ostertag:2011}
Eric Ostertag.
\newblock {\em Mono- and Multivariable Control and Estimation}.
\newblock Springer, 2011.

\bibitem{Qin:2003}
S.~Joe Qin and Thomas~A. Badgwell.
\newblock A survey of industrial model predictive control technology.
\newblock {\em Control Engineering Practice}, 11(7):733--764, 2003.
\newblock \href {http://dx.doi.org/10.1016/S0967-0661(02)00186-7}
  {\path{doi:10.1016/S0967-0661(02)00186-7}}.

\bibitem{Radke:2006}
Aaron Radke and Zhiqiang Gao.
\newblock A survey of state and disturbance observers for practitioners.
\newblock In {\em Proceedings of the 2006 American Control Conference}, pages
  5183--5188, 2006.
\newblock \href {http://dx.doi.org/10.1109/ACC.2006.1657545}
  {\path{doi:10.1109/ACC.2006.1657545}}.

\bibitem{Su:2005}
Y.~X. Su, C.~H. Zheng, and B.~Y. Duan.
\newblock Automatic disturbances rejection controller for precise motion
  control of permanent-magnet synchronous motors.
\newblock {\em IEEE Transactions on Industrial Electronics}, 52(3):814--823,
  2005.
\newblock \href {http://dx.doi.org/10.1109/TIE.2005.847583}
  {\path{doi:10.1109/TIE.2005.847583}}.

\bibitem{Sun:2005}
Bosheng Sun and Zhiqiang Gao.
\newblock A {DSP}-based active disturbance rejection control design for a
  1-{kW} {H}-bridge {DC}-{DC} power converter.
\newblock {\em IEEE Transactions on Industrial Electronics}, 52(5):1271--1277,
  2005.
\newblock \href {http://dx.doi.org/10.1109/TIE.2005.855679}
  {\path{doi:10.1109/TIE.2005.855679}}.

\bibitem{Vincent:2011}
John Vincent, Dan Morris, Nathan Usher, Zhiqiang Gao, Shen Zhao, Achille
  Nicoletti, and Qinling Zheng.
\newblock On active disturbance rejection based control design for
  superconducting {RF} cavities.
\newblock {\em Nuclear Instruments and Methods in Physics Research A},
  643(1):11--16, 2011.
\newblock \href {http://dx.doi.org/10.1016/j.nima.2011.04.033}
  {\path{doi:10.1016/j.nima.2011.04.033}}.

\bibitem{Zheng:2010}
Qing Zheng and Zhiqiang Gao.
\newblock On practical applications of active disturbance rejection control.
\newblock In {\em Proceedings of the 29th Chinese Control Conference}, pages
  6095--6100, 2010.

\end{thebibliography}
\end{document}